\documentclass[lettersize,journal]{IEEEtran}
\usepackage{amsmath,amsfonts}
\usepackage{amssymb}
\usepackage{algorithm}
\usepackage{algorithmic}
\usepackage{array}
\usepackage[caption=false,font=normalsize,labelfont=sf,textfont=sf]{subfig}
\usepackage{textcomp}
\usepackage{stfloats}
\usepackage{url}
\usepackage{verbatim}
\usepackage{newtxtext}
\usepackage{newtxmath}
\usepackage{cite}
\usepackage{multirow}
\usepackage{makecell}
\usepackage{color}
\usepackage{balance}
\usepackage{bm}

\newcommand{\LoS}{\cellcolor{blue!12}}
\newcommand{\OLoS}{\cellcolor{orange!15}}
\renewcommand{\arraystretch}{1.25}
\usepackage[table]{xcolor}
\usepackage{graphicx}
\begin{document}

\title{Channel Measurements and Characterization with Phase Drift Compensation for Outdoor 330--360~GHz MIMO Communications}

\author{Tian~Qiu,~Taihao~Zhang,~Cunhua~Pan,~\IEEEmembership{Senior~Member,~IEEE},~Hong~Ren,~\IEEEmembership{Member,~IEEE},~Yongchao~He,~Chenzhou~Lin,~Jiaming~Zeng,~Bingchang~Hua,~and~Jiangzhou~Wang,~\IEEEmembership{Life Fellow,~IEEE}

\thanks{Tian Qiu, Taihao Zhang, Cunhua Pan, Hong Ren, Yongchao He, Chenzhou Lin, Jiaming Zeng and Jiangzhou Wang are with the National Mobile Communications Research Laboratory, Southeast University, Nanjing, China (e-mail: tianqiu, taihao, cpan, hren, heyongchao, chenzhoulin, jmzeng, j.z.wang@seu.edu.cn).}
\thanks{Bingchang Hua is with Purple Mountain Laboratories, Nanjing, Jiangsu 211111, China (e-mail: huabingchang@pmlabs.com.cn).}
\thanks{Corresponding author: Cunhua Pan.}}

\maketitle

\begin{abstract}
This paper presents an outdoor channel measurement campaign at 330--360~GHz using a \(128\times4\) virtual antenna array (VAA) for multiple-input multiple-output (MIMO) channel sounding. The measured links are classified as line-of-sight (LoS) or obstructed-LoS (OLoS) to characterize outdoor terahertz (THz) propagation under different obstruction conditions. During the measurement process, the stationarity of the outdoor environment is carefully verified, and phase drift (PD) with an approximately linear trend is observed. Then, we propose a PD-aware space-alternating generalized expectation-maximization (SAGE) algorithm, which significantly improves both delay resolution and channel parameter estimation accuracy. Based on the processed measurement data, we characterize key channel properties, including the power delay profile, path loss, shadow fading, delay spread, angular spread, Rician K-factor, as well as their cumulative distribution functions and correlation characteristics. In addition, MIMO-specific properties, including near-field effects, spatial non-stationarity and cluster birth--death property, are analyzed.
\end{abstract}

\section{Introduction}

\IEEEPARstart{T}{erahertz} (THz) communication is widely recognized as a key enabling technology for sixth-generation (6G) wireless communications, offering abundant spectrum resources to alleviate the spectrum scarcity and capacity limitations of current wireless systems~\cite{intro1,intro2}. Channel measurements and characterization are essential for understanding propagation mechanisms and provide a fundamental basis for system design and deployment~\cite{intro3,intro4}. Therefore, to fully exploit the potential of the THz band, channel measurement campaigns in representative scenarios are necessary~\cite{intro15,intro14}.

Since channel characteristics strongly depend on the operating frequency range, propagation environment, and deployment scenario, dedicated channel measurement campaigns must be conducted in representative scenarios of interest. Indoor THz channel measurements have been extensively investigated. In~\cite{intro6}, sub-THz channel characterization and modeling were performed for indoor industrial scenarios at 142~GHz in four factories, where 82 transmitter--receiver (Tx--RX) locations under both line-of-sight (LoS) and non-LoS (NLoS) conditions were selected and more than 75,000 spatial and temporal channel impulse responses (CIRs) were collected. In~\cite{intro7}, indoor channel measurement campaigns were conducted at 201--209~GHz, covering four communication scenarios, including a meeting room, cubicle area, hallway, and an NLoS case. In~\cite{indoor}, an indoor channel measurement campaign at 260--400~GHz was reported using a 128-element virtual multi-antenna linear array, thereby improving the understanding of THz multi-antenna channel characteristics. 

Although these studies have advanced the understanding of indoor THz propagation, their findings do not necessarily extend to outdoor environments, where propagation geometries, scatterer distributions, and obstruction conditions differ. Therefore, dedicated outdoor channel measurement campaigns remain indispensable for characterizing these environment-dependent propagation effects and establishing an empirical basis for outdoor THz MIMO channel modeling.

\subsection{Related Work}

\subsubsection{THz-band Outdoor Channel Measurement}

Recently, considerable efforts have been devoted to outdoor THz channel measurements. At 142~GHz, NYU WIRELESS conducted a series of outdoor urban microcell (UMi) measurement studies for sub-THz channel characterization~\cite{outdoor1,outdoor2,outdoor3,outdoor8}. Specifically,~\cite{outdoor1} investigated the spatial autocorrelation properties of shadow fading (SF), delay spread (DS), and angular spread (AS). In~\cite{outdoor2}, outdoor omnidirectional and directional path loss (PL) models were provided for both LoS and NLoS scenarios, along with an analysis of foliage loss. In~\cite{outdoor3}, the channel spatial statistics, including the number of spatial clusters and the cluster power distribution, were studied. Building upon these measurement efforts,~\cite{outdoor8} summarized extensive measurement results and statistical analyses to formulate a multiband empirical three-dimensional statistical channel model (SCM) for outdoor urban open square and street scenarios.~\cite{outdoor4} conducted wideband channel measurements in an outdoor street canyon environment around the 300~GHz band. The analysis of PL and root-mean-square (RMS) DS showed that outdoor environments exhibit more severe large-scale fading and greater delay dispersion. In~\cite{outdoor7}, wideband channel measurements were conducted along an L-shaped university campus street at 306--321~GHz and 356--371~GHz. The results verified the sparsity of multipath components (MPCs) at THz frequencies and showed smaller power spreads in both the temporal and spatial domains in the THz band.~\cite{outdoor5} and~\cite{outdoor6} conducted channel measurements and statistical modeling at 145--146~GHz for urban device-to-device (D2D) and UMi environments, respectively, using vector network analyzer (VNA)-based channel sounders. These studies investigated key channel parameters and provided statistical models that enable a more realistic and detailed performance assessment of THz communication systems. Nevertheless, outdoor THz-band multiple-input multiple-output (MIMO) channel measurements and characterization remain insufficiently investigated.

\subsubsection{Stationarity Verification}

Many existing measurement studies assume that the channel environment is quasi-static or semi-static~\cite{outdoor1,SAGE3}, without explicitly verifying its stationarity. However, in virtual antenna array (VAA) measurements, the extended acquisition duration may introduce phase drift (PD) errors due to environmental changes and hardware drift. Theoretical analyses and experimental studies have demonstrated the presence of PD and its adverse effects. In~\cite{PD1}, the authors pointed out that VAA sounders are susceptible to phase-noise-induced measurement errors, and demonstrated that local oscillator (LO) impairments can lead to significant errors. In~\cite{PD5}, an optical fiber-based VNA sounder exhibited phase variations associated with cable bending and thermal changes. At sub-THz frequencies,~\cite{PD4} measured peak-to-peak phase drifts of \(21^\circ\) and \(60^\circ\) in the WR05 and WR03 bands, respectively. In~\cite{PD3}, reference antenna measurements were used to estimate and correct PD errors during post-processing, and an iterative low-complexity CLEAN algorithm was adopted for MPC extraction. Nevertheless, stationarity verification and PD compensation for VNA-based VAA MIMO measurements in the THz band remain relatively underexplored.

\subsubsection{Measurement Data Processing Methods}

The space-alternating generalized expectation-maximization (SAGE) algorithm was originally proposed as a generalization of the classical expectation-maximization (EM) algorithm~\cite{SAGE1}. By alternating among multiple hidden-data spaces and updating only a subset of the parameter vector in each iteration, SAGE can accommodate smoothness penalties and achieve faster convergence than conventional EM algorithms. In~\cite{SAGE2}, the SAGE algorithm was applied to mobile radio channel measurements under both synthetic and real-world conditions, demonstrating its effectiveness in the high-resolution extraction of MPC parameters. Owing to its high resolution and iterative refinement capability, SAGE has been widely adopted for wireless channel parameter estimation~\cite{SAGE3}.

\subsection{Contributions}

To address the aforementioned gaps, we conduct outdoor MIMO channel measurements in the 330--360~GHz band using a VAA configuration. The main contributions of this paper are summarized as follows:
\begin{itemize}
	\item Stationarity verification reveals PD in outdoor VAA measurements, with its dominant trend characterized by linear fitting. Motivated by this observation, we develop a PD-aware SAGE algorithm that incorporates PD refinement into iterative multipath parameter extraction. Simulations demonstrate lower channel frequency response (CFR) reconstruction normalized mean square error (NMSE) than that
	obtained with multiple signal classification (MUSIC) and PD-unaware SAGE. Measurements at TRx1 further demonstrate improved reconstruction of a weak delayed component, with the mean reconstruction NMSE reduced from \(0.0390\) to \(0.0164\) relative to PD-unaware SAGE.
	
    \item Based on the measured data, we characterize the power delay profile (PDP), PL, SF, DS, AS, and Rician K-factor (KF), and consolidate the location-specific statistics and fitted model parameters into a measurement-based reference for outdoor THz channel characterization. For DS, AS, and KF, five candidate distributions are compared using the Kolmogorov--Smirnov (KS) distance, with their parameter estimates and fitting results systematically tabulated. The comparison favors lognormal (LN) models for DS, AS, and KF within the analyzed dataset.
    
    \item In addition, CIR correlations and MIMO-specific properties, including near-field effects, spatial non-stationarity (SnS) and cluster birth--death property, are analyzed. Sliding-subarray angle-of-departure (AoD) estimates, path-power profiles, and delay-domain observations reveal position-dependent local angles, path- and link-dependent power variations, and spatially localized cluster visibility. These complementary observations highlight the importance of accounting for element-dependent path strength and visibility in THz MIMO channel models.

    \item We classify the measured TRx links as LoS or obstructed-LoS (OLoS) and compare their channel characteristics. The OLoS links exhibit larger departure-angle spreads in all four examined neighboring pairs and lower spatial CIR correlations in the two evaluated pairs. These findings highlight obstruction-related differences in the angular dispersion and spatial structure of the measured outdoor THz channels.
\end{itemize}

\subsection{Organization and Notation}

The remainder of this paper is organized as follows. Section~\ref{measurement} describes the measurement environment and system setup. Section~\ref{PDSAGE} presents the stationarity verification and the proposed \(\mathrm{PD}\)-aware SAGE algorithm. The outdoor MIMO channel characteristics and modeling results are analyzed in Section~\ref{results}. Finally, Section~\ref{conclusion} concludes the paper.

\textit{Notation:} Boldface lowercase and uppercase letters denote vectors and matrices, respectively. The \(m\)-th entry of \(\mathbf x\) and the \((m,n)\)-th entry of \(\mathbf X\) are denoted by \([\mathbf x]_m\) and
\([\mathbf X]_{m,n}\), respectively. For a matrix \(\mathbf X\), \(\mathbf X^*\), \(\mathbf X^{\mathrm T}\), and \(\mathbf X^{\mathrm H}\) denote its complex conjugate, transpose, and conjugate transpose, respectively. The modulus of a scalar is denoted by \(|\cdot|\), the Euclidean norm of vector \(\mathbf{x}\) is denoted by \(\|\mathbf{x}\|_2\), and the Frobenius norm of matrix \(\mathbf{X}\) is denoted by \(\|\mathbf{X}\|_F\). The vectorization operator \(\mathrm{vec}(\mathbf{X})\) stacks the columns of \(\mathbf{X}\) into a column vector, whereas \(\mathrm{mat}(\mathbf{X})\) reshapes \(\mathbf{X}\) into a matrix of specified dimensions following column-wise ordering. The expectation operator is denoted by \(\mathbb{E}\left[\cdot\right]\). Additionally, the Kronecker product and transposed Khatri-Rao product between two matrices \(\mathbf{X}\) and \(\mathbf{Y}\) are denoted by \(\mathbf{X} \otimes \mathbf{Y}\) and \(\mathbf{X} \bullet \mathbf{Y}\), respectively.  

\begin{figure*}[b]
    \centering
    \subfloat[Measurement environment.]{
        \includegraphics[width=0.48\textwidth]{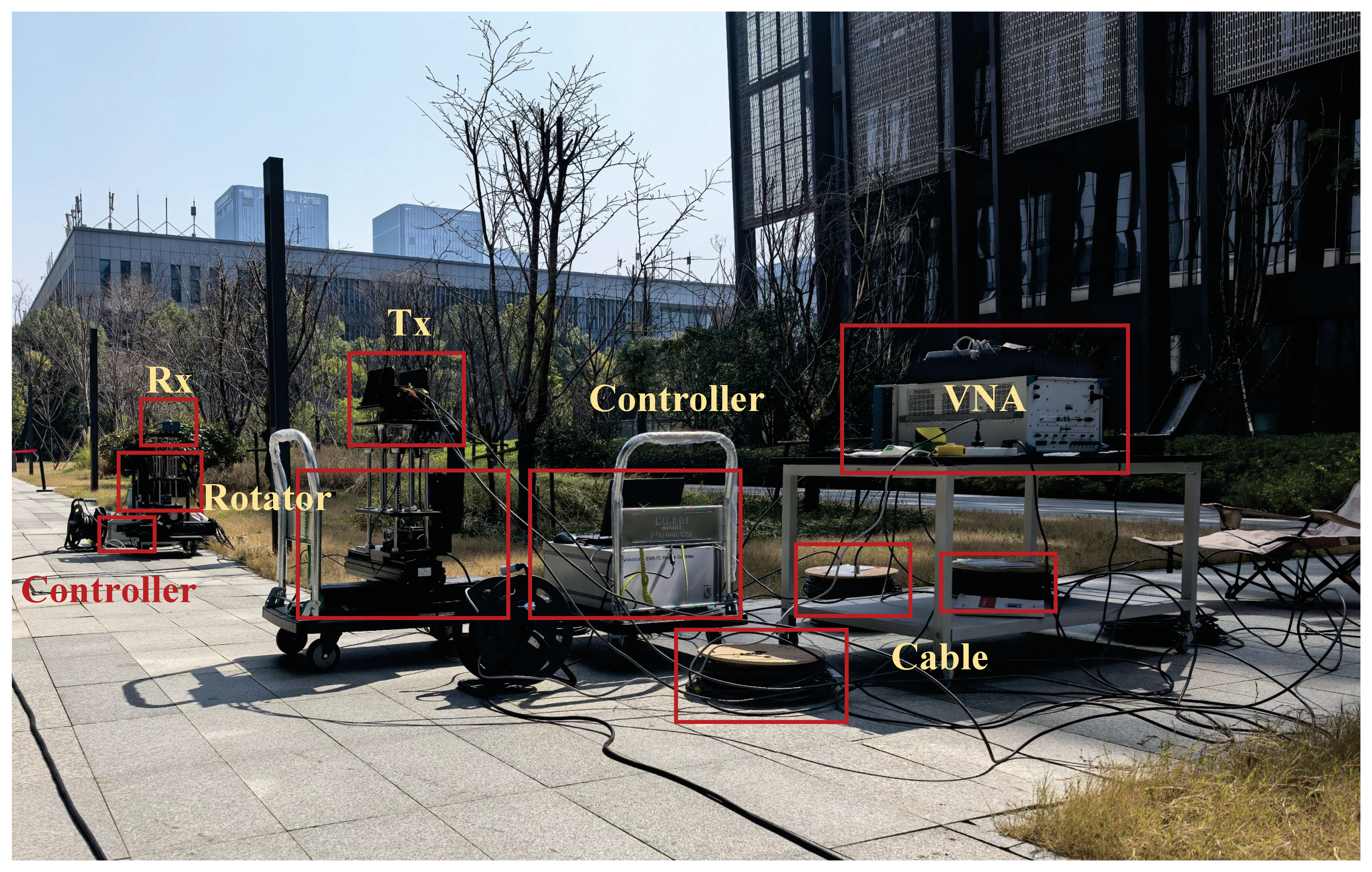}
        \label{fig_photo}
    }
    \hfill
    \subfloat[Measurement layout.]{
        \includegraphics[width=0.48\textwidth]{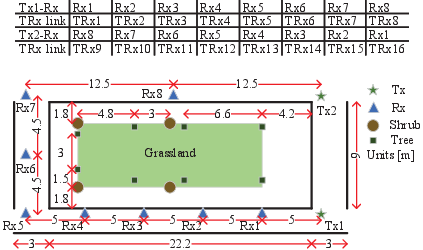}
        \label{fig_layout}
    }
    \caption{Outdoor THz MIMO measurement campaign: (a) measurement setup and (b) layout.}
    \label{fig_environment}
\end{figure*}

\section{Outdoor THz MIMO Channel Measurement}\label{measurement}

In this section, the outdoor measurement campaign conducted in the 330--360~GHz band is presented. To capture the spatial characteristics of outdoor THz propagation, a \(128\times4\) VAA-based MIMO configuration is employed. In the following, we first describe the outdoor measurement environment and the classification of TRx links, and then introduce the VNA-based measurement system and its key configuration parameters.

\subsection{Measurement Environment}\label{environment}

The 330--360~GHz outdoor measurement campaign is conducted in a rectangular street area at Purple Mountain Laboratories (PML), Nanjing, as shown in Fig.~\ref{fig_environment}. Specifically, Fig.~\ref{fig_environment}\subref{fig_photo} presents a photograph of the measurement environment, while Fig.~\ref{fig_environment}\subref{fig_layout} illustrates the corresponding layout, including the Tx and Rx locations as well as the positions of the main objects in the environment. For notational simplicity, the links between Tx location 1 and Rx locations 1-8 are denoted as TRx locations 1-8, while those from Tx location 2 to Rx locations 8-1 are denoted as TRx locations 9-16. The grass in the central area is approximately \(0.6~\mathrm{m}\) tall, while the shrubs and trees are approximately \(1.4~\mathrm{m}\) and \(3~\mathrm{m}\) tall, respectively. Based on a joint assessment of the measurement layout, vegetation distribution, and measured channel characteristics, TRx1-TRx5, TRx8-TRx10, and TRx14-TRx16, corresponding to 11 locations, are classified as LoS cases, whereas TRx6-TRx7 and TRx11-TRx13, corresponding to 5 locations, are classified as OLoS cases.

To characterize the spatial structure of the outdoor THz channel, we employ a mechanically synthesized VAA, following established approaches to high-frequency multi-antenna channel sounding~\cite{VAA,indoor,sns,VAA2,VAA3}. High-precision motorized linear stages sequentially position the Tx and Rx antennas along predefined linear trajectories, synthesizing uniform linear arrays (ULAs) with \(N_T=128\) and \(N_R=4\) virtual elements, respectively. The spatial sampling intervals are set to \(d_T=d_R=\lambda_c/2\), where \(\lambda_c\) denotes the wavelength at the center frequency.

The asymmetric array configuration is selected to prioritize departure-side spatial characterization while constraining the duration of sequential acquisition. The 128-element Tx array provides an aperture of \(D=(N_T-1)d_T=63.5\lambda_c\), supporting AoD estimation and the characterization of spatial channel variations across the aperture~\cite{VAA3}. At the Rx, four virtual elements provide receive-side spatial observations without the acquisition overhead associated with a comparably large array. This configuration requires 512 Tx--Rx element-pair measurements per link, reducing the acquisition time relative to configurations with large arrays at both ends and consequently limiting exposure to temporal variations during outdoor measurements~\cite{VAA}.

\subsection{Measurement System Setup}\label{system}

\begin{table}[H]
\centering
\caption{Measurement System Setup}
\label{tab_measurement_system}
\rowcolors{2}{gray!15}{white}
\resizebox{1\columnwidth}{!}{
\begin{tabular}{c c c c}
\hline
\rowcolor{gray!30}
\textbf{Parameter} & \textbf{Value} & \textbf{Parameter} & \textbf{Value} \\
\hline
Start frequency & \(330~\mathrm{GHz}\) & End frequency & \(360~\mathrm{GHz}\) \\
Time domain resolution & \(33.33~\mathrm{ps}\) & Bandwidth & \(30~\mathrm{GHz}\) \\
Path length resolution & \(1~\mathrm{cm}\) & IFBW & \(1~\mathrm{kHz}\) \\
Maximum excess delay & \(166.67~\mathrm{ns}\) & Sweeping points & \(5001\) \\
Maximum path length & \(49.97~\mathrm{m}\) & Sweeping interval & \(6~\mathrm{MHz}\) \\
Antenna gain at Tx & \(25~\mathrm{dBi}\) & HPBW of Tx & \(10^{\circ}\) \\
Antenna gain at Rx & \(25~\mathrm{dBi}\) & HPBW of Rx & \(10^{\circ}\) \\
Average noise floor & \(-145~\mathrm{dBm}\) & Test power & \(0.5~\mathrm{mW}\) \\
\hline
\end{tabular}
}
\end{table}

The outdoor MIMO channel measurements in the THz band are conducted using a VNA-based frequency-domain measurement system operating in the 330--360~GHz band. The detailed operating principle of the measurement system can be found in~\cite{indoor}. As shown in Fig.~\ref{fig_environment}\subref{fig_photo}, in the transmitter module, the VNA generates a radio frequency (RF) signal spanning \(12.22\)-\(13.33~\mathrm{GHz}\), which is up-converted by a \(\times\) 27 frequency multiplier (FM). Meanwhile, an LO generates a signal over \(13.75\)-\(15~\mathrm{GHz}\), which is up-converted by a \(\times\) 24 FM. The two resulting signals are then mixed to generate a \(279~\mathrm{MHz}\) intermediate-frequency (IF) reference signal. An analogous process is performed in the receiver module to generate the corresponding test signal. The key system configuration parameters and measurement setup are summarized in Table~\ref{tab_measurement_system}.

\section{Stationarity Verification and PD-Aware SAGE Algorithm}\label{PDSAGE}

Most existing channel measurement studies assume that the propagation environment remains quasi-static during the measurement process~\cite{outdoor1,SAGE3}; however, this assumption is seldom verified experimentally, especially for VAA-based THz MIMO measurements, where the sequential antenna movement leads to an extended acquisition duration. Therefore, before extracting channel characteristics, it is necessary to examine the temporal stationarity of the measurement environment. In this section, we first verify the stationarity of the measurement environment and reveal the existence of an approximately linear PD. Based on this observation, we further propose a PD-aware SAGE algorithm that incorporates PD compensation into the iterative MPC extraction process, thereby improving the reliability and resolution of channel parameter estimation.
\subsection{Phase Drift Characterization}\label{linearPD}

To assess environmental stationarity under fixed propagation geometry, repeated frequency sweeps are performed at selected TRx links using the system configuration in Section~\ref{system}. The Tx and Rx antennas are held stationary at the first Tx and Rx virtual element positions with their boresight directions aligned. For each link, the phase at a selected frequency point is unwrapped across successive sweeps and referenced to the first record.

\begin{figure}
    \centering
    \includegraphics[width=3.4in]{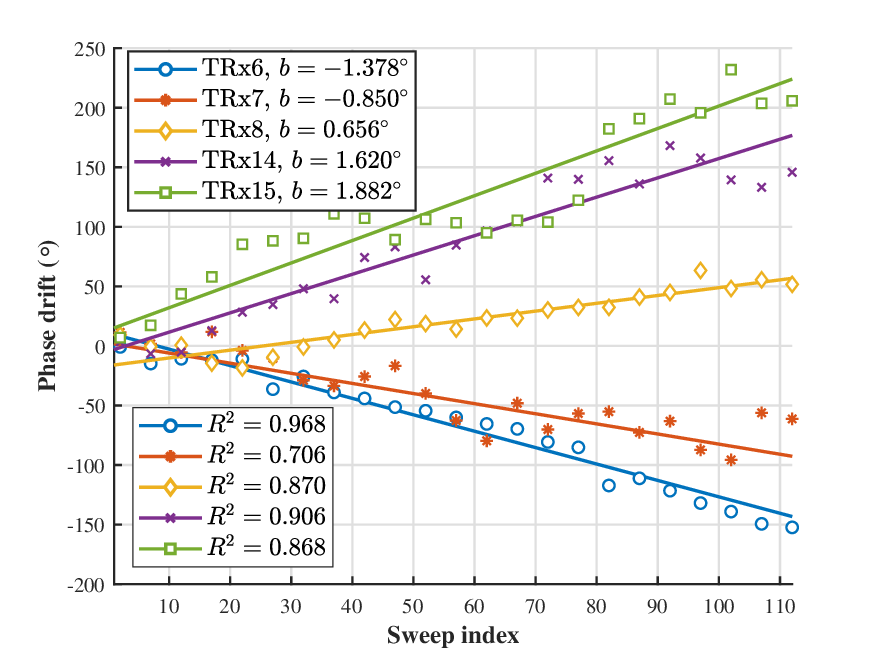}
    \caption{Relative phase drift and linear fits across successive frequency sweeps at selected TRx links.}
\label{fig_pdyanzheng}
\end{figure}

The measured amplitude remains nearly constant across repeated sweeps, whereas the phase exhibits an approximately linear drift. Accordingly, a linear model is fitted to the measured relative phase to quantify its evolution over the acquisition sequence. The results are presented in Fig.~\ref{fig_pdyanzheng}, where the markers and solid lines represent the measured samples and linear fits, respectively. The horizontal axis denotes successive complete frequency sweeps rather than elapsed time, since the interval between records need not equal the nominal VNA sweep duration. Thus, the fitted slope characterizes the phase increment per sweep rather than a drift rate per unit time. The goodness of fit is quantified by the coefficient of determination,
\begin{align}
    R^2 = 1-\frac{\sum_{i=1}^{N}\left(\varphi_i-\widehat{\varphi}_i\right)^2}{\sum_{i=1}^{N}\left(\varphi_i-\bar{\varphi}\right)^2},
    \label{R2}
\end{align}
where \(\varphi_i\), \(\widehat{\varphi}_i\), and \(\bar{\varphi}\) denote the measured relative phase, its fitted value, and its sample mean, respectively, and \(N\) is the number of sweeps. The average \(R^2\) of \(0.864\) indicates that the linear model captures the dominant phase evolution over the examined acquisition sequences. The fitted slope, denoted by \(b\), is expressed in degrees per sweep.

The observed PD may result from thermally induced variations in the electrical length of the cables and the phase response of the measurement chain~\cite{PD4,PD5}. Differences in thermal conditions and cable configuration between measurements may account for the different slopes across TRx links. These slopes are therefore treated as run-specific drift indicators rather than intrinsic channel parameters. Under a locally linear drift assumption, the observed range of \(b\) provides empirical guidance for selecting a coarse search interval for the PD parameter \(\phi\).

\subsection{PD-Aware SAGE Algorithm}

The VNA measures and records the scattering parameters. The effects of the system frequency response are removed
through amplitude and phase calibration. The calibrated frequency-domain response between the \(t\)-th Tx antenna element and the \(r\)-th Rx antenna element at the \(k\)-th frequency point is
\begin{align}
    {H}^{(r,t)}[k] 
    &= \frac{S_{21}^{(r,t)}[k]}{H_{\mathrm{RF}}^{\mathrm{Tx}}[k] H_{\mathrm{device}}[k] H_{\mathrm{RF}}^{\mathrm{Rx}}[k]},
    \label{S21}
\end{align}
where \(1\leq k\leq K\). \(H_{\mathrm{RF}}^{\mathrm{Tx}}[k]\) and \(H_{\mathrm{RF}}^{\mathrm{Rx}}[k]\) denote the frequency responses of the RF front ends at the Tx and Rx, respectively, and \(H_{\mathrm{device}}[k]\) accounts for the frequency response contributions from the VNA and cables. Finally, \({H}^{(r,t)}[k]\) represents the calibrated CFR. By arranging \({H}^{(r,t)}[k]\) over all Rx--Tx antenna element pairs, the MIMO CFR matrix under the VAA configuration is obtained as
\begin{align}
    \mathbf{H}[k] 
    = \sqrt{\frac{N_R N_T}{L}} 
    \sum_{l=1}^{L} \alpha_{l} 
    e^{-j2\pi k \Delta f \tau_l} 
    \mathbf{a}_{N_R}\left(\psi_{l}\right) 
    \mathbf{a}_{N_T}^{\mathrm{H}}\left(\omega_{l}\right),
    \label{CFRMIMO}
\end{align}
where \(\alpha_l\), \(\tau_l\), \(\psi_l\), and \(\omega_l\) denote the complex path gain, delay, angle of arrival (AoA), and AoD of the \(l\)-th MPC, respectively. \(\Delta f\) is the frequency sweep interval. \(\mathbf{a}_{N_R}(\psi_l)\in\mathbb{C}^{N_R\times 1}\) and \(\mathbf{a}_{N_T}(\omega_l)\in\mathbb{C}^{N_T\times 1}\) denote the array response vectors (ARVs) at the Rx and Tx, respectively, and are given by

\begin{subequations}
\begin{align}
    \mathbf{a}_{N_R}(\psi_l) 
    &= \frac{1}{\sqrt{N_R}} e^{-\mathrm{j}\frac{2\pi}{\lambda_c}d_R\sin(\psi_l)
    \left[0, 1, \cdots, N_R-1 \right]^{\mathrm{T}}}, 
    \label{ARV1} \\
    \mathbf{a}_{N_T}(\omega_l) 
    &= \frac{1}{\sqrt{N_T}} 
    e^{-\mathrm{j}\frac{2\pi}{\lambda_c}d_T\sin(\omega_l)
    \left[0, 1, \cdots, N_T-1 \right]^{\mathrm{T}}},
    \label{ARV2}
\end{align}
\end{subequations}
where \(\lambda_c\), \(d_R\), and \(d_T\) denote the carrier wavelength, Rx antenna spacing, and Tx antenna spacing, respectively.

Next, we incorporate the linear PD observed in Section~\ref{linearPD} into the channel model. The PD-aware CFR matrix, denoted by \(\mathbf{H}_{\mathrm{pd}}[k]= \mathbf{H}[k] e^{\mathrm{j}k\phi}\), is expressed as
\begin{align}
    \mathbf{H}_{\mathrm{pd}}[k] 
    =  
    \sum_{l=1}^{L} \tilde{\alpha}_l 
    e^{-\mathrm{j}2\pi k \Delta f \left(\tau_l-\frac{\phi}{2\pi \Delta f}\right)}
    \mathbf{a}_{N_R}\left(\psi_l\right)
    \mathbf{a}_{N_T}^{\mathrm{H}}\left(\omega_l\right),
    \label{pd}
\end{align}
where \(\tilde{\alpha}_l \triangleq \sqrt{\frac{N_R N_T}{L}} \alpha_l\) denotes the equivalent path gain~\footnote{In practical channel measurement processing, the normalization factor \(\sqrt{\frac{N_R N_T}{L}}\) is omitted. In addition, the carrier-frequency-dependent phase term associated with \(f_c\), which should be included in Eq.~\eqref{CFRMIMO}, is also absorbed into the equivalent complex path gain during parameter extraction and does not affect the subsequent analysis.}. As can be observed from Eq.~\eqref{pd}, the effective delay becomes \(\tilde{\tau}_l \triangleq \tau_l-\frac{\phi}{2\pi\Delta f}\). Therefore, failure to properly account for PD will result in biased delay estimates, which may consequently degrade the estimation accuracy of other channel parameters.

Moreover, the CFR in Eq.~\eqref{pd} can be expressed in a compact matrix form as
\begin{align}
    \mathbf{H}_{\mathrm{pd}}[k] = \mathbf{A}_{N_R} \mathbf{\Lambda}[k] \mathbf{A}_{N_T}^{\mathrm{H}},
    \label{compact}
\end{align}
where \(\mathbf{A}_{N_R} = \left[ \mathbf{a}_{N_R} \left( \psi_{1} \right), \cdots, \mathbf{a}_{N_R} \left( \psi_{L} \right) \right] \in \mathbb{C}^{N_R\times L}\) and \(\mathbf{A}_{N_T} = \left[ \mathbf{a}_{N_T} \left( \omega_{1} \right), \cdots, \mathbf{a}_{N_T} \left( \omega_{L} \right) \right] \in \mathbb{C}^{N_T\times L}\) denote the AoA and AoD steering matrices, respectively, and \(\mathbf{\Lambda}[k] = \mathrm{Diag} \left( \tilde{\alpha}_1 e^{-j 2\pi k \Delta f \tilde{\tau}_1}, \cdots, \tilde{\alpha}_L e^{-j 2\pi k \Delta f \tilde{\tau}_L} \right) \in \mathbb{C}^{L\times L}\) denotes the frequency-dependent equivalent complex gain matrix.

The SAGE algorithm is widely used for the offline processing of large channel measurement datasets~\cite{SAGE3}. Unlike the conventional EM algorithm, which relies on a single complete-data space and updates all parameters simultaneously, SAGE alternates among multiple admissible hidden-data spaces and updates only a subset of the parameter vector in each iteration by maximizing the objective function. In practical channel parameter estimation, this maximization step can be implemented using high-resolution parameter estimation techniques. Building upon this framework, we propose a PD-aware SAGE algorithm to jointly refine the MPC parameters and compensate for measurement-induced PD. The proposed algorithm follows the SAGE framework and iteratively updates each MPC using an orthogonal matching pursuit (OMP)-type procedure.

In the expectation-step (\textbf{E-step}), the admissible hidden data associated with the \(l\)-th MPC are constructed by subtracting the currently reconstructed contributions of all other MPCs from the received matrix, i.e.,
\begin{align}
    \widehat{\mathbf{H}}_{\mathrm{pd}}^{l}[k]=\mathbf{H}_{\mathrm{pd}}[k]-\sum_{l^{\prime}=1,l^{\prime}\ne l}^{\widehat{L}}\widehat{\mathbf{H}}_{\mathrm{pd}}^{l^{\prime}}[k],
    \quad 1\leq k\leq K.
    \label{Estep}
\end{align}
In the maximization steps (\textbf{M-steps}), \(\operatorname{OMP}(\mathbf{Y},\mathbf{A},J)\) denotes an
OMP-based recovery procedure with observation matrix \(\mathbf Y\), sensing dictionary \(\mathbf A\), and
sparsity level \(J\). In the following, we detail the construction of the observation matrices and sensing matrices for estimating each MPC parameter. The complete PD-aware SAGE procedure is summarized in Algorithm~\ref{alg1}. The proposed PD-aware SAGE framework is not restricted to the present measurement scenario. It can be adapted to other configurations by characterizing the PD and adjusting the sensing dictionary accordingly.

Let the PD-aware delay response vector of the \(l\)-th MPC be defined as 
\begin{align}
    \mathbf{t}_{\mathrm{pd}}(\tilde{\tau}_l)=\left[e^{-\mathrm{j}2\pi\Delta f\tilde{\tau}_l},\cdots,e^{-\mathrm{j}2\pi K\Delta f\tilde{\tau}_l}\right]^{\mathrm{T}}\in\mathbb{C}^{K\times 1}.
\end{align}
The \(l\)-th PD-aware delay observation matrix is constructed by vectorizing the \(l\)-th path component at each frequency point and concatenating the resulting vectors over frequency.
\begin{align}
    \mathbf{Y}_{\tau_l}
    & \triangleq
    \left[
    \mathrm{vec}(\widehat{\mathbf{H}}_{\mathrm{pd}}^{l}[1]),
    \cdots,
    \mathrm{vec}(\widehat{\mathbf{H}}_{\mathrm{pd}}^{l}[K])
    \right]^{\mathrm{T}} \notag \\
    & \approx \tilde{\alpha}_l \mathbf{t}_{\mathrm{pd}}(\tilde{\tau}_l)
    \left[
    \mathbf{a}_{N_T}^{\mathrm{*}}(\omega_l)
    \otimes
    \mathbf{a}_{N_R}(\psi_l)
    \right]^{\mathrm{T}}
    \in\mathbb{C}^{K\times N_RN_T}.
    \label{Ytau}
\end{align}
To estimate the delay with high precision while reducing computational complexity, the dominant delay window is identified from the CIR obtained by inverse fast Fourier transform (IFFT). Specifically, for the \((r,t)\)-th Rx--Tx element pair, the equivalent single-input single-output (SISO) frequency response of the \(l\)-th MPC is given by
\begin{align}
    (\widehat{\mathbf{h}}_{\mathrm{pd}}^{l})^{(r,t)}
    & \triangleq
    \left[
    [\widehat{\mathbf{H}}_{\mathrm{pd}}^{l}[1]]_{r,t},
    \cdots,
    [\widehat{\mathbf{H}}_{\mathrm{pd}}^{l}[K]]_{r,t}
    \right]^{\mathrm{T}} \notag \\
    & \approx \tilde{\alpha}_l
    \left[\mathbf{a}_{N_R}(\psi_l)\right]_r
    \left[\mathbf{a}_{N_T}^{*}(\omega_l)\right]_t
    \mathbf{t}_{\mathrm{pd}}(\tilde{\tau}_l)
    \in\mathbb{C}^{K\times1}.
    \label{SISO}
\end{align}
Thus, the average PDP (APDP) of the \(l\)-th MPC, denoted by \(\mathrm{APDP}_{l}\), is obtained by averaging the PDPs reconstructed from the \(l\)-th MPC over all Rx--Tx antenna element pairs. Specifically, the reconstructed CIR for the \((r,t)\)-th antenna element pair is
\begin{align}
(\widehat{\mathbf{c}}_{\mathrm{pd}}^{l})^{(r,t)}=\operatorname{IFFT}((\widehat{\mathbf{h}}_{\mathrm{pd}}^{l})^{(r,t)})\in\mathbb{C}^{K\times 1}.
\end{align}
Then, \(\mathrm{APDP}_{l}\) is computed as 
\begin{align}
    \mathrm{APDP}_{l}
    =
    \frac{1}{N_RN_T}
    \sum_{r=1}^{N_R}
    \sum_{t=1}^{N_T}
    \left|
    (\widehat{\mathbf{c}}_{\mathrm{pd}}^{l})^{(r,t)}
    \right|^2.
    \label{APDPl}
\end{align}
The dominant delay window is identified from the peak of \(\mathrm{APDP}_l\), and a local delay grid \(\boldsymbol{\tau}_l=[\tau_{l,1},\cdots,\tau_{l,D_{\tau_l}}]\in\mathbb{R}^{1\times D_{\tau_l}}\) is constructed within this window, where \(D_{\tau_l}\) denotes the delay dictionary size. The corresponding delay sensing matrix is
\begin{align}
\mathbf A_{\tau_l}
=
\exp_{\circ}\!\left[
-\mathrm j2\pi
\left(\mathbf f\otimes\boldsymbol{\tau}_l\right)
\right]
\in\mathbb C^{K\times D_{\tau_l}},
\end{align}
where \(\exp_{\circ}(\cdot)\) denotes the elementwise exponential and
\(\mathbf f=[\Delta f,\ldots,K\Delta f]^{\mathrm T}\) denotes the frequency sampling vector. Constructing \(\mathbf{A}_{\tau_l}\) only within the dominant delay window improves delay resolution while reducing the search complexity~\footnote{In measurement data processing, the measured PDP can be directly obtained from the data. Therefore, the delay window range can be manually fine-tuned to achieve higher-precision delay estimation.}.

After the coarse delay estimate \(\widehat{\tau}_l^c\) is obtained, a PD-aware sensing matrix is constructed
to jointly refine the delay and PD estimates. Specifically, a locally refined delay sensing matrix \(\mathbf{A}_{\widehat{\tau}_l}\) is constructed around the coarse estimate \(\widehat{\tau}_l^{\mathrm{c}}\). Let \(\boldsymbol{\phi}=[\phi_1,\ldots,\phi_{D_\phi}]\in\mathbb{R}^{1\times D_\phi}\) denote the PD search grid, where \(D_\phi\) is the dictionary size. The search range is determined using the prior
information obtained in Section~\ref{linearPD}. The corresponding PD sensing matrix is defined as

\begin{align}
\mathbf{A}_{\mathrm{pd}} = \exp_{\circ}\!\left[
\mathrm j
\left(\frac{\mathbf f}{\Delta f}\otimes\boldsymbol{\phi}\right)
\right] \in\mathbb{C}^{K\times {D_{\phi}}}. 
\end{align}
Thus, the joint sensing matrix is constructed as \(\mathbf{A}_{\widehat{\tau}_l}\bullet\mathbf{A}_{\mathrm{pd}}\).

Once the refined delay estimate \(\widehat{\tau}_l^r\) and the PD estimate \(\widehat{\phi}\) have been obtained, the corresponding PD-aware delay atom is reconstructed as \(\mathbf t_{\mathrm{pd}}(\widehat{\widetilde{\tau}}_l)\). For angle-domain parameter estimation, the delay-domain influence is eliminated by projecting \(\mathbf{Y}_{\tau_l}\) onto the estimated PD-aware delay response vector. The \(l\)-th AoA observation matrix is then constructed by reshaping the projected spatial-domain response into an \(N_R\times N_T\) matrix as
\begin{align}
    \mathbf{Y}_{R_l}
    \triangleq
    \operatorname{mat}
    \left(
    \frac{\mathbf{t}_{\mathrm{pd}}^{\mathrm{H}}(\widehat{\tilde{\tau}}_l)
    \mathbf{Y}_{\tau_l}}{K}
    \right) \approx
    \mathbf{a}_{N_R}(\psi_l)
    \left(
    \tilde{\alpha}_l
    \mathbf{a}_{N_T}^{\mathrm{H}}(\omega_l)
    \right).
    \label{YAoA}
\end{align}
To estimate the \(l\)-th AoA \(\psi_l\), the Rx-side sensing dictionary is constructed as 
\begin{align}
\mathbf{A}_{R}=\left[\mathbf{a}_{N_R}(-1),\mathbf{a}_{N_R}(-1+2/D_R),\cdots,\mathbf{a}_{N_R}(1-2/D_R)\right],
\end{align}
where \(D_R\geq N_R\) denotes the dictionary size and the grid resolution is \(2/D_R\). Similarly, the \(l\)-th AoD observation matrix is constructed as 
\begin{align}
\mathbf{Y}_{T_l} \triangleq \left(\mathbf{Y}_{R}^{l}\right)^{\mathrm{H}} \approx \mathbf{a}_{N_T}(\omega_l) \left( \tilde{\alpha}_l^{*} \mathbf{a}_{N_R}^{\mathrm{H}}(\psi_l) \right) \in \mathbb{C}^{N_T\times N_R}.
\end{align}
The \(l\)-th AoD \(\omega_l\) can be estimated in the same manner using the Tx-side sensing dictionary \(\mathbf{A}_{T}\)~\footnote{The order of delay-domain and angle-domain parameter estimation is not fixed and can be adjusted according to their relative influence on the overall estimation accuracy. For example, \(\mathbf{Y}_{R_l}\) can be constructed before delay estimation as \(\mathbf{Y}_{R_l} \triangleq \left[ \widehat{\mathbf{H}}_{\mathrm{pd}}^l[1], \cdots, \widehat{\mathbf{H}}_{\mathrm{pd}}^l[K] \right] \approx \mathbf{a}_{N_R}(\psi_l) \left( \tilde{\alpha}_l \mathbf{t}_{\mathrm{pd}}^{\mathrm{T}}(\tilde{\tau}_l) \otimes \mathbf{a}_{N_T}^{\mathrm{H}}(\omega_l) \right) \in \mathbb{C}^{N_R \times K N_T}\).}.

Then, by exploiting the approximate orthogonality of the large-scale array response vectors, the equivalent complex path gain \(\alpha_l\) is estimated as \(\widehat{\alpha}_l = \mathbf{a}_{N_R}(\widehat{\psi}_l)^{\mathrm{H}} \mathbf{Y}_{R_l} \mathbf{a}_{N_T}(\widehat{\omega}_l)\).

\begin{algorithm}[t]
\caption{PD-Aware SAGE Algorithm}
\label{alg1}
\begin{algorithmic}[1]
\REQUIRE \(\mathbf{H}_{\mathrm{pd}}[k]\), \(1\leq k\leq K\); stopping threshold \(\epsilon\); maximum number of iterations \(\delta_{\max}\); sensing matrices \(\{\mathbf{A}_{\mathrm{pd}}, \mathbf{A}_{R}, \mathbf{A}_{T}\}\); estimated number of MPCs \(\widehat{L}\).
\STATE \textbf{Initialization:} Set \(\mathrm{NMSE}^{(0)}=0\) and iteration index \(\delta=1\). Initialize \(\{\widehat{\mathbf{H}}_{\mathrm{pd}}^{l}[k]\}_{l=1}^{\widehat{L}}\) and compute \(\mathrm{NMSE}^{(1)}\).
\WHILE{\(\left|\mathrm{NMSE}^{(\delta)}-\mathrm{NMSE}^{(\delta-1)}\right|>\epsilon\) and \(\delta<\delta_{\max}\)}
\STATE \(\delta \leftarrow \delta+1\)
\FOR{\(1\leq l\leq \widehat{L}\)}
\STATE \textbf{E-step:} \(\widehat{\mathbf{H}}_{\mathrm{pd}}^{l}[k]=\mathbf{H}_{\mathrm{pd}}[k]-\sum_{l^{\prime}=1,l^{\prime}\ne l}^{\widehat{L}}\widehat{\mathbf{H}}_{\mathrm{pd}}^{l^{\prime}}[k]\). 
\STATE Obtain \(\mathbf{Y}_{\tau_l}\) via Eq. \eqref{Ytau}.
\STATE Obtain \(\widehat{\mathbf{h}}_{\mathrm{pd}}^{l,(r,t)}\) via Eq. \eqref{SISO}.
\STATE Obtain \(\boldsymbol{\tau_{l}}\) from \(\mathrm{APDP}_{l}\). 
\STATE Construct the coarse delay sensing matrix \(\mathbf{A}_{\tau_l}\).
\STATE \textbf{M-step:} \(\widehat{\tau}_l^c=\operatorname{OMP} (\mathbf{Y}_{\tau_l},\mathbf{A}_{\tau_l},1)\). 
\STATE Construct the locally refined delay sensing matrix \(\mathbf{A}_{\widehat{\tau}_l}\).
\STATE \textbf{M-step:} \((\widehat{\tau}_l^r,\widehat{\phi})=\operatorname{OMP}(\mathbf{Y}_{\tau_l},\mathbf{A}_{\widehat{\tau}_l} \bullet \mathbf{A}_{\mathrm{pd}},1)\).
\STATE Obtain \(\mathbf{Y}_{R_l}\) via Eq. \eqref{YAoA}.
\STATE \textbf{M-step:} \(\widehat{\psi}_l=\operatorname{OMP}(\mathbf{Y}_{R_l},\mathbf{A}_{R},1)\).
\STATE Obtain \(\mathbf{Y}_{T_l}=\left(\mathbf{Y}_{R}^{l}\right)^{\mathrm{H}}\).
\STATE \textbf{M-step:} \(\widehat{\omega}_l=\operatorname{OMP}(\mathbf{Y}_{T_l},\mathbf{A}_{T},1)\).
\STATE \textbf{M-step:} \(\widehat{\alpha}_l = \mathbf{a}_{N_R}(\widehat{\psi}_l)^{\mathrm{H}} \mathbf{Y}_{R_l} \mathbf{a}_{N_T}(\widehat{\omega}_l)\).
\STATE Update the estimated \(l\)-th PD-aware MPC \(\widehat{\mathbf{H}}_{\mathrm{pd}}^{l}[k]\).
\ENDFOR
\STATE Update the estimated CFR as \(\widehat{\mathbf{H}}_{\mathrm{pd}}[k]=\sum_{l=1}^{\widehat{L}}\widehat{\mathbf{H}}_{\mathrm{pd}}^{l}[k]\).
\STATE Compute \(\mathrm{NMSE}^{(\delta)}\).
\ENDWHILE
\ENSURE \(\{\widehat{\alpha}_l,\widehat{\tau}_l,\widehat{\psi}_l,\widehat{\omega}_l\}_{l=1}^{\widehat{L}}\), \(\widehat{\phi}\), \(\widehat{\mathbf{H}}_{\mathrm{pd}}[k]\), \(1\leq k\leq K\).
\end{algorithmic}
\end{algorithm}

In the proposed algorithm, \(\widehat{L}\) is predefined to be sufficiently large to ensure that all significant propagation paths in the measurement data can be captured. This overcomplete setting is commonly adopted for measured channels~\cite{SAGE3}, where the exact number of MPCs is unknown in advance. However, because \(\widehat{L}\) is intentionally selected to be large, some estimated components may correspond to weak or negligible paths. Although PD affects all MPCs, \(\phi\) is not estimated uniformly after all \(\widehat{L}\) path components have been updated, since including these weak components in a global PD estimation may introduce additional errors and degrade the estimation accuracy. Instead, the PD estimation is incorporated into the path-wise refinement process, where the delay and PD are jointly estimated based on the corresponding dominant hidden-data component.

Moreover, proper initialization can accelerate convergence. This can be achieved by performing the M-steps once or by applying the MUSIC algorithm~\cite{SAGE2} before iterative refinement. For example, solving the OMP problem associated with Eq.~\eqref{compact} with sparsity level \(\widehat{L}\) yields the initial estimates \(\{\widehat{\psi}_l\}_{l=1}^{\widehat{L}}\). 

\begin{figure*}[t]
    \centering
    \subfloat[Simulation-based validation, with \(\widehat{L}=L\), \(\delta_{\max}=5\), and \(\epsilon=10^{-6}\).]{
        \includegraphics[width=0.48\textwidth]{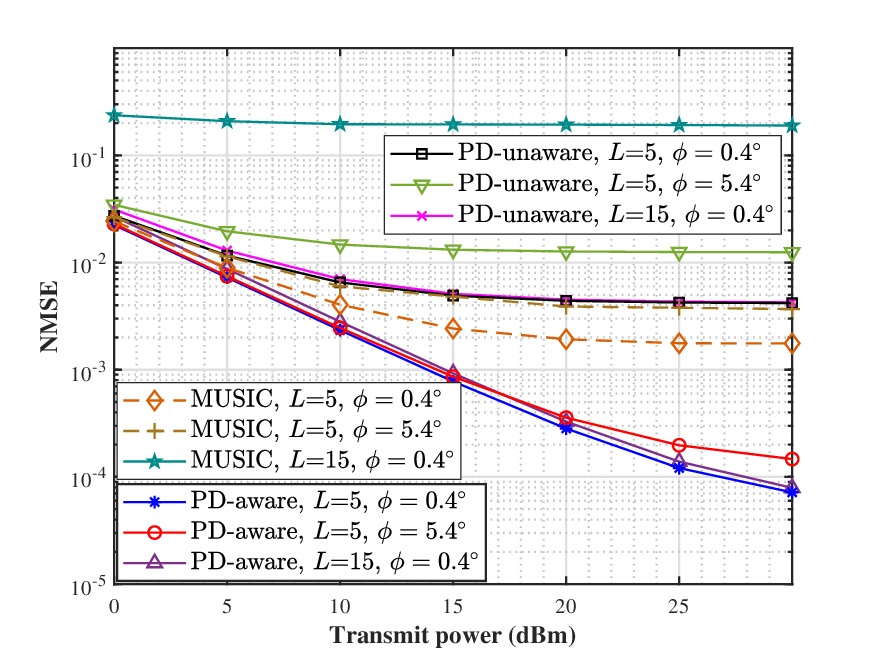}
        \label{fig_pdsimulation}
    }
    \hfill
    \subfloat[Measurement-based validation, with \(\widehat{L}=100\), \(\delta_{\max}=5\), and \(\epsilon=10^{-4}\).]{
        \includegraphics[width=0.48\textwidth]{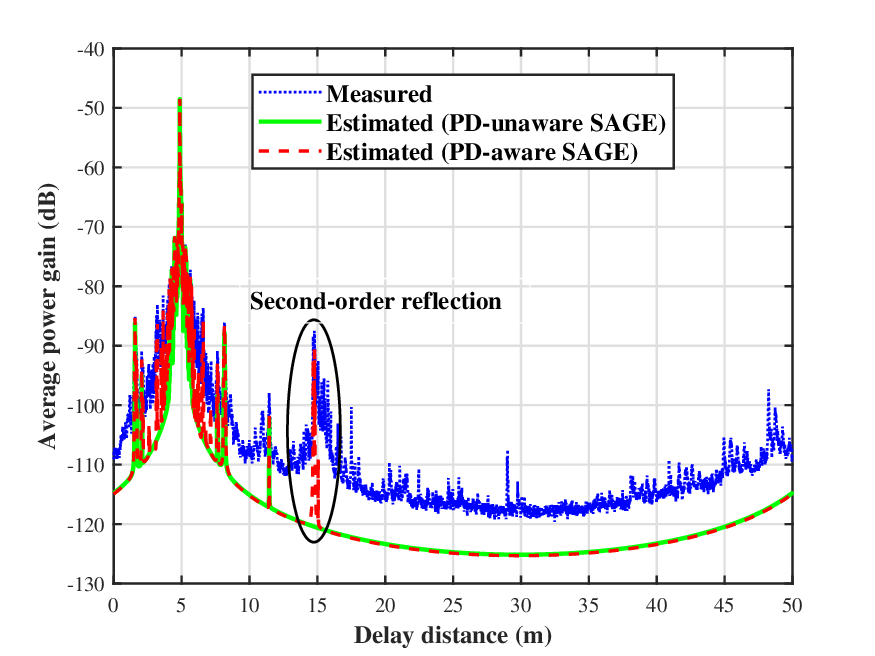}
        \label{fig_pdmeasured}
    }
    \caption{Simulation and measurement evaluation of PD-aware SAGE:
    	(a) NMSE versus transmit power and
    	(b) measured and reconstructed APDPs at TRx1.}
    \label{fig_pd}
\end{figure*}

Fig.~\ref{fig_pd}\subref{fig_pdsimulation} illustrates the simulation-based validation of the proposed PD-aware SAGE algorithm. In the simulation, the received signal is modeled as \(\mathbf{Y}[k]=\sqrt{p}\mathbf{H}_{\mathrm{pd}}[k]+\mathbf{N}[k]\), where \(p\) denotes the transmit power and \(\mathbf{N}[k]\) denotes additive white Gaussian noise (AWGN). The Tx and Rx are equipped with ULAs consisting of \(N_T=64\) and \(N_R=64\) antenna elements, respectively. The antenna spacings are set to \(d_T=d_R=\lambda_{\mathrm{c}}/2\), where \(\lambda_{\mathrm{c}}\) denotes the wavelength corresponding to the center frequency. The complex path gains are modeled as \(\alpha_l\sim\mathcal{CN}(0,10^{-3}d_{\mathrm{TR}}^{-2.2})\), where \(d_{\mathrm{TR}}=50~\mathrm{m}\) denotes the Tx--Rx distance. The noise power is set to \(\sigma^2=-80~\mathrm{dBm}\). The dictionary sizes are set to \(D_R=N_R\) and \(D_T=N_T\), and \(\psi_l\) and \(\omega_l\) are randomly generated from the corresponding angular grids. In addition, the propagation distance corresponding to each delay is uniformly generated within \(\mathcal{U}[5,15]~\mathrm{m}\). The PD \(\phi\) is uniformly sampled from \([0.4^{\circ},5^{\circ}]\). The remaining parameters are set consistently with those of the measurement system. 

MUSIC and PD-unaware SAGE are used as baselines. The PD-unaware SAGE baseline omits the joint PD-aware sensing matrix and the explicit \(\phi\) update. Since the ground-truth CFR \(\mathbf{H}_{\mathrm{pd}}[k]\) is unavailable for actual measurement data, the NMSE is computed as follows
\begin{align}
    \mathrm{NMSE} = \mathbb{E}\left\{ \frac{\sum_{k=1}^{K}\left\|\widehat{\mathbf{H}}_{\mathrm{pd}}[k]-\mathbf{Y}[k]\right\|_{{F}}^{2}}{\sum_{k=1}^{K}\left\|\mathbf{Y}[k]\right\|_{{F}}^{2}} \right\}.
    \label{NMSE}
\end{align}
As observed from the figure, the proposed PD-aware SAGE algorithm achieves a lower NMSE than both MUSIC and the PD-unaware baseline for all configurations considered. Increasing the number of MPCs \(L\) increases the number of unknown parameters and degrades estimation accuracy, with MUSIC exhibiting the most pronounced degradation. A larger PD \(\phi\) also increases the NMSE, particularly for the PD-unaware baseline, since uncompensated PD introduces an effective delay shift that biases delay estimates and can affect the selection of the dominant signal window. At high transmit powers, the NMSEs of both baselines tend to saturate, whereas that of the proposed algorithm continues to decrease over the considered power range. These results demonstrate the superior estimation accuracy of the proposed algorithm in scenarios with stronger PD or a larger number of MPCs.
 
The proposed PD-aware algorithm is then applied to the measured data\footnote{For notational simplicity, the estimation marker \(\widehat{\cdot}\) and the subscript \(\mathrm{pd}\) are omitted in the following sections unless otherwise specified.}. For the \((r,t)\)-th Rx--Tx antenna element pair, the reconstructed CIR is obtained by applying the \( \mathrm{IFFT} \) to the sum of all estimated MPC CFRs, i.e., 
\begin{align}
{\mathbf{c}}^{(r,t)}=\operatorname{IFFT} \left(\sum_{l=1}^{{L}} ({\mathbf{h}}^{l})^{(r,t)}\right)\in\mathbb{C}^{K\times1}.
\end{align}
The corresponding PDP is given by the squared magnitude of the CIR. The APDP is then computed by averaging the PDPs over all Rx--Tx antenna element pairs as
\begin{align}
    \mathrm{APDP}
    =
    \frac{1}{N_RN_T}
    \sum_{r=1}^{N_R}
    \sum_{t=1}^{N_T}
    \left|
    {\mathbf{c}}^{(r,t)}
    \right|^2.
    \label{APDP}
\end{align}
Fig.~\ref{fig_pd}\subref{fig_pdmeasured} presents the measurement-based validation of the proposed PD-aware SAGE algorithm using the data measured at the TRx1 location. As observed, the PD-aware algorithm improves the delay-domain resolution. Specifically, both the PD-aware and PD-unaware algorithms can identify the LoS path at approximately \(5~\mathrm{m}\), whereas only the PD-aware algorithm successfully resolves the second-order reflection path caused by the measurement devices at approximately \(15~\mathrm{m}\). The mean NMSE values over all Rx--Tx antenna element pairs are calculated as \(\bar{\mathrm{NMSE}}_{\mathrm{aware}}=\frac{1}{N_RN_T} \sum_{r=1}^{N_R} \sum_{t=1}^{N_T}\mathrm{NMSE}_{\mathrm{aware}}^{(r,t)}=0.0164\) and \(\bar{\mathrm{NMSE}}_{\mathrm{unaware}}=0.0390\), respectively. In addition, the relative NMSE reduction is first computed for each Rx--Tx antenna element pair and then averaged over all pairs, yielding an average improvement of \(61.571\%\). 

\section{Channel Characterization and Modeling}\label{results}

Following the stationarity assessment and PD-aware processing in Section~\ref{PDSAGE}, this section presents a comprehensive characterization and modeling of the measured outdoor THz-band MIMO channels. Based on the processed channel responses and extracted MPC parameters, both large-scale and small-scale channel characteristics are analyzed. In addition, MIMO-specific propagation phenomena are investigated to provide a more complete understanding of outdoor THz MIMO propagation. A comparative analysis between LoS and OLoS cases is also conducted to reveal the impact of vegetation obstruction on the measured channel characteristics. Table~\ref{tab_results} summarizes the measurement and fitting results for the selected channel parameters, where the blue-highlighted TRx locations correspond to LoS cases and the yellow-highlighted TRx locations correspond to OLoS cases.

\begin{table*}[t]
\centering
\caption{Measured Channel Statistics and Fitted Model Parameters}
\large
\label{tab_results}
\renewcommand{\arraystretch}{1.3} 
\resizebox{\textwidth}{!}{
\begin{tabular}{>{\columncolor{gray!30}\centering\arraybackslash}m{0.45cm}|c|*{15}{c|}c}
\hline
\rowcolor{gray!30}
\multicolumn{2}{c|}{\textbf{TRx location}}
& \LoS 1 & \LoS 2 & \LoS 3 & \LoS 4 & \LoS 5 & \OLoS 6 & \OLoS 7 & \LoS 8 
& \LoS 9 & \LoS 10 & \OLoS 11 & \OLoS 12 & \OLoS 13 & \LoS 14 & \LoS 15 & \LoS 16 \\
\hline

& PL (dB)
& 95.037 & 101.450 & 104.418 & 107.774 & 108.901 & 108.853 & 108.234 & 105.467 & 102.502 & 108.853 & 108.746 & 117.594 & 111.250 & 107.049 & 104.021 & 105.669 \\
\cline{2-18}

& CI model 
& \multicolumn{16}{c}{\(n=1.900\)} \\
\cline{2-18}

\multirow{-3}{*}{\textbf{PL}}
& FI model 
& \multicolumn{16}{c}{\(\alpha=2.108,\ \beta=80.573~\mathrm{dB}\)} \\

\hline

& CI model (dB)
& -1.228 & -0.769 & -0.823 & -0.366 & -0.924 & -1.172 & -2.285 & -0.697 & -1.274 & -0.809 & -1.144 & 7.141 & 2.406 & -0.369 & -1.082 & 2.327 \\
\cline{2-18}

& FI model (dB)
& -0.023 & -0.215 & -0.599 & -0.458 & -1.201 & -1.471 & -2.638 & -0.573 & -0.890 & -1.068 & -1.428 & 6.796 & 2.236 & -0.382 & -0.843 & 2.758 \\
\cline{2-18}

\multirow{-3}{*}{\textbf{SF}}
& \(\sigma_{\mathrm{SF}}\) (dB)
& \multicolumn{16}{c}{
\(\sigma_{\mathrm{SF}}^{\mathrm{CI}}=2.207,\ \sigma_{\mathrm{SF}}^{\mathrm{FI}}=2.167\quad\)
} \\
\hline

& \(\sigma_{\tau}\) (ns)
& 0.606 & 0.755 & 0.362 & 1.023 & 0.244 & 0.227 & 0.665 & 0.656 & 0.245 & 0.446 & 0.549 & 0.610 & 0.330 & 0.405 & 0.425 & 0.456 \\
\cline{2-18}

\multirow{-2}{*}{\textbf{DS}}
& LN (\(\log_{10}(\mathrm{DS}/1\mathrm{ns})\))
& \multicolumn{16}{c}{
\(\mu_{\mathrm{DS}}=-0.348,\ \sigma_{\mathrm{DS}}=0.197\)
} \\
\hline

& \(\sigma_{\omega}\) (\(^\circ\))
& 3.768 & 2.351 & 2.893 & 1.972 & 2.418 & 4.353 & 3.187 & 1.732 & 2.850 & 4.287 & 5.651 & 13.542 & 5.658 & 4.309 & 5.505 & - \\
\cline{2-18}

\multirow{-2}{*}{\textbf{AS}}
& LN (\(\log_{10}(\mathrm{AS}/1^{\circ})\))
& \multicolumn{16}{c}{
\(\mu_{\mathrm{AS}}=0.553,\ \sigma_{\mathrm{AS}}=0.261\)
} \\
\hline

& \(\mu_{\kappa}\) (dB)
& 14.528 & 7.730 & 15.876 & 13.035 & 18.655 & 19.728 & 9.275 & 8.933 & 21.117 & 14.097 & 12.019 & 11.393 & 19.658 & 16.037 & 15.305 & 14.279 \\
\cline{2-18}

& \(\sigma_{\kappa}\) (dB)
& 5.290 & 1.608 & 2.889 & 0.934 & 0.986 & 1.052 & 0.917 & 0.692 & 3.204 & 1.111 & 1.372 & 1.529 & 3.789 & 1.459 & 1.859 & 1.102 \\
\cline{2-18}

\multirow{-3}{*}{\textbf{KF}}
& LN (\(\log_{10}(\kappa)\))
& \multicolumn{16}{c}{
\(\mu_{\mathrm{KF}}=1.448,\ \sigma_{\mathrm{KF}}=0.449\)
} \\
\hline
\end{tabular}
}
\end{table*}

\subsection{Average Power Delay Profile}

As an example, the measured and estimated APDPs at the TRx1 location are shown in Fig.~\ref{fig_pd}\subref{fig_pdmeasured}. The estimated results agree well with the measured data, indicating that most MPCs are accurately extracted, although some weak, long-delay paths are not detected. A strong LoS path is observed, with an average power gain of \(-48.528~\mathrm{dB}\), whereas the second-order reflection path has an average power gain of \(-90.675~\mathrm{dB}\). Moreover, the spurious peaks near the path (SPNP) phenomenon has been discussed in detail in our previous work~\cite{indoor}.

\begin{figure*}[t]
    \centering
    \subfloat[Path loss.]{
        \includegraphics[width=0.48\textwidth]{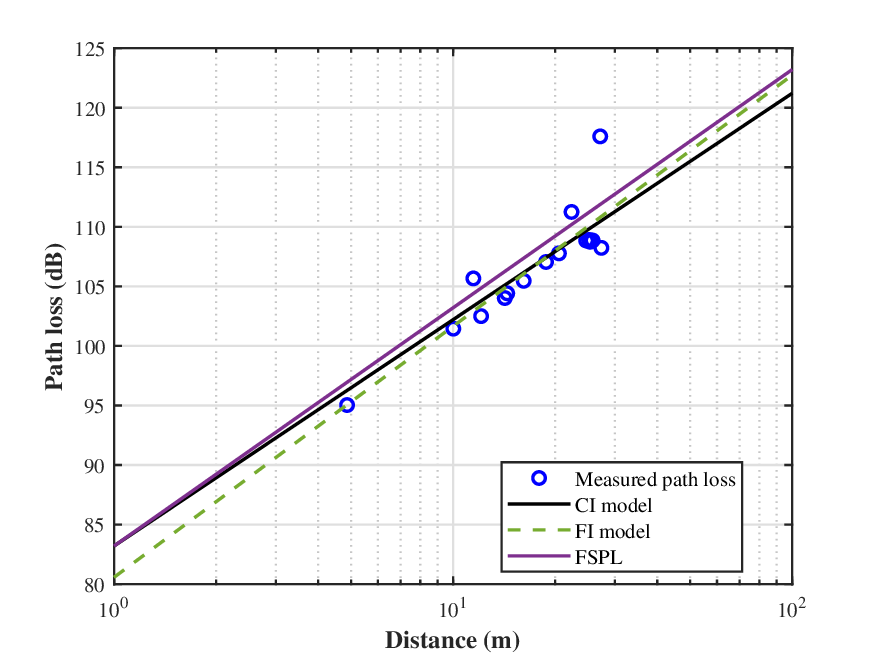}
        \label{fig_pl}
    }
    \hfill
    \subfloat[Shadow fading.]{
        \includegraphics[width=0.48\textwidth]{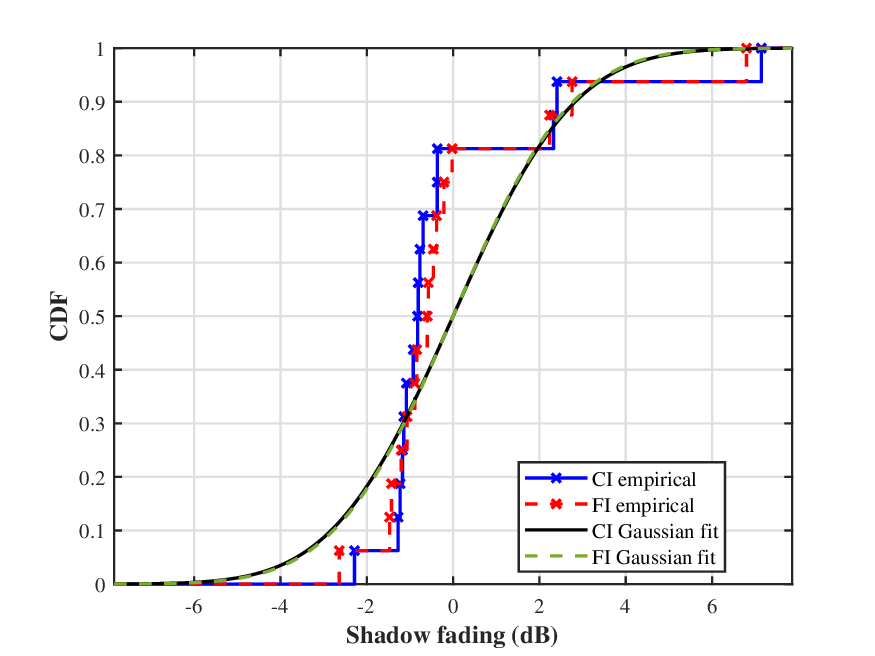}
        \label{fig_sf}
    }
    \caption{Path loss and shadow fading of the measured outdoor THz links:
    	(a) measured PL and CI/FI model fits;
    	(b) empirical CDFs of SF and Gaussian fits.}
    \label{fig_plsf}
\end{figure*}

\subsection{Path Loss and Shadow Fading}

Based on Friis' transmission equation, the PL between the Tx and Rx antenna ports is expressed in dB as
\begin{align}
    \mathrm{PL}
    =
    10\log_{10}\frac{P_t}{P_r}
    =
    -10\log_{10}
    \left[
    G_tG_r
    \left(
    \frac{\lambda_c}{4\pi d_{\mathrm{TR}}}
    \right)^2
    h
    \right],
    \label{PLmeas}
\end{align}
where \(P_t=0.5~\mathrm{mW}\) and \(P_r\) denote the transmitted and received powers, respectively. \(G_t\) and \(G_r\) denote the linear antenna power gains, each corresponding to \(25~\mathrm{dBi}\), and \(h = \sum_{k=1}^{K}\mathrm{APDP}[k]\). The close-in (CI) and floating-intercept (FI) models are adopted to characterize the PL. The CI model is given by
\begin{align}
    \mathrm{PL}^{\mathrm{CI}}
    =
    10 n \log_{10}\left(\frac{d_{\mathrm{TR}}}{d_0}\right)
    +
    \mathrm{FSPL}(f_c,d_0)
    +
    X^{\mathrm{CI}}_{\sigma},
    \label{PLCI}
\end{align}
where \(n\) is the path loss exponent (PLE), and \(d_0\) is the reference distance, which is set to \(1~\mathrm{m}\) in this scenario. \(X_{\sigma}^{\mathrm{CI}}\) denotes the SF term, modeled as a zero-mean Gaussian variable with standard deviation \(\sigma_{\mathrm{SF}}^{\mathrm{CI}}\). The free-space path loss (FSPL) is given by
\begin{align}
\mathrm{FSPL}(f_c,d) = 20\log_{10}\left(\frac{4\pi d f_c}{c}\right),
\end{align}
where \(c\) denotes the speed of light. Moreover, the \(\mathrm{FI}\) path loss model is expressed as
\begin{align}
    \mathrm{PL}^{\mathrm{FI}}
    =
    10 \alpha \log_{10}\left(d_{\mathrm{TR}}\right)
    +
    \beta
    +
    X^{\mathrm{FI}}_{\sigma},
    \label{PLFI}
\end{align}
where \(\alpha\) and \(\beta\) denote the linear slope and floating intercept, respectively. \(X_{\sigma}^{\mathrm{FI}}\) denotes the SF term, modeled as a zero-mean Gaussian variable with standard deviation \(\sigma_{\mathrm{SF}}^{\mathrm{FI}}\). The corresponding measurement and fitting results are presented in Fig.~\ref{fig_plsf} and Table~\ref{tab_results}.

Fig.~\ref{fig_plsf}\subref{fig_pl} compares the measured PL with the CI and FI models fitted to the pooled measurements from all 16 links. The CI model yields \(n=1.900\), close to the free-space exponent of two and the LoS values reported for millimeter-wave (mmWave) UMi street-canyon channels~\cite{pathloss}. The FI model, with \(\alpha=2.108\) and \(\beta=80.573~\mathrm{dB}\), provides similar PL estimates over the measured distance range, indicating limited improvement from allowing a free intercept. These pooled fits characterize the overall distance-dependent attenuation, while the SF residuals describe location-specific departures from the fitted trend.

Fig.~\ref{fig_plsf}\subref{fig_sf} shows similar empirical CDFs for the CI and FI residuals, with pooled SF standard deviations of \(2.207\) and \(2.167~\mathrm{dB}\), respectively. Both values are below the \(4~\mathrm{dB}\) UMi street-canyon LoS reference in 3GPP TR~38.901~\cite{3gpp38901}. The location-specific residuals nevertheless reveal appreciable differences. Some OLoS links, including TRx6, TRx7, and TRx11, exhibit
SF residuals comparable to those of the LoS links. In contrast, TRx12 exhibits the largest positive residual under both models, at \(7.141~\mathrm{dB}\) for CI and \(6.796~\mathrm{dB}\) for FI. This contrast is consistent with location-dependent attenuation associated with vegetation obstruction, indicating that a small pooled SF standard deviation does not preclude substantial excess attenuation at individual locations.

\subsection{Delay and Angular Spreads}

\begin{figure*}[t]
    \centering
    \subfloat[Delay spread.]{
        \includegraphics[width=0.48\textwidth]{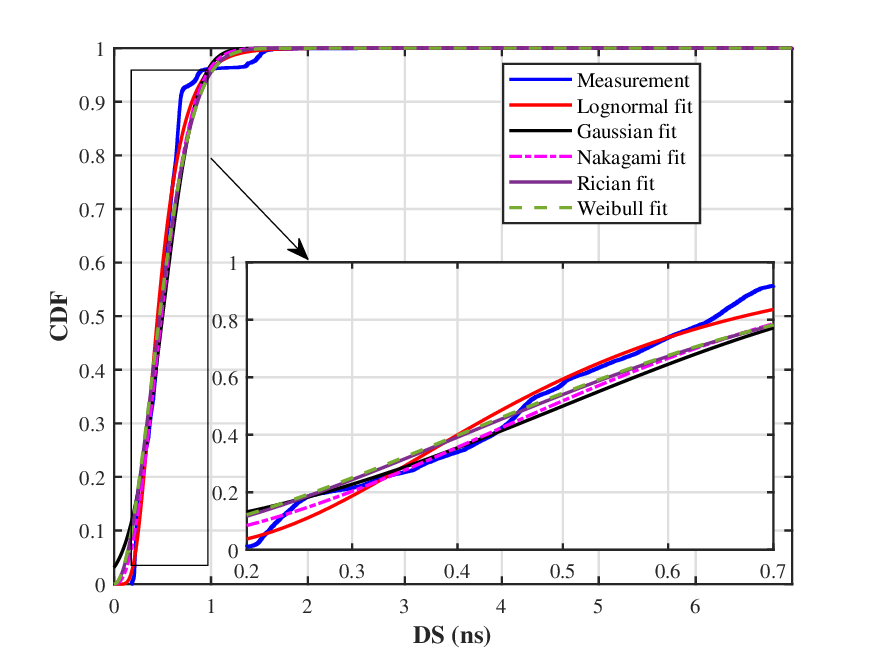}
        \label{fig_ds}
    }
    \hfill
    \subfloat[Angular spread.]{
        \includegraphics[width=0.48\textwidth]{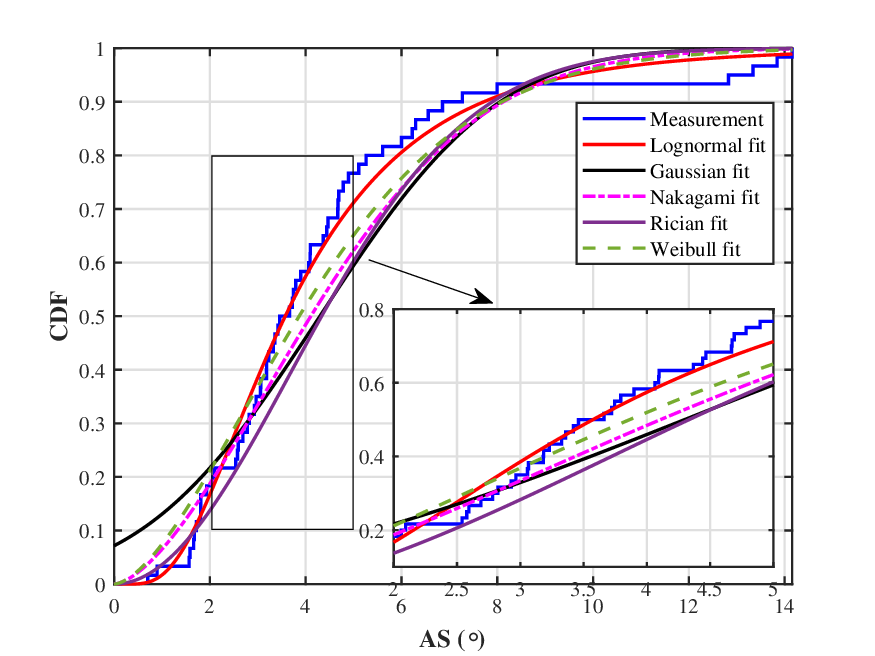}
        \label{fig_as}
    }
    \caption{Empirical and fitted CDFs of
    	(a) RMS delay spread over all 16 links and (b) RMS departure-angle spread over TRx1--TRx15.}
    \label{fig_dsas}
\end{figure*}

The RMS DS is widely used to characterize delay dispersion in the channel and is calculated as
\begin{align}
    \sigma_{\tau}
    =\sqrt{\frac{\sum_{l=1}^{L} |\alpha_l|^2 \tau_l^2}{\sum_{l=1}^{L} |\alpha_l|^2}-\left(\frac{\sum_{l=1}^{L} |\alpha_l|^2 \tau_l}{\sum_{l=1}^{L} |\alpha_l|^2}\right)^2}.
    \label{DS}
\end{align}
Similarly, the RMS AS characterizes the dispersion of channel power across different angular directions and is given by
\begin{align}
    \sigma_{\omega}=\sqrt{\frac{\sum_{l=1}^{L} |\alpha_l|^2 \omega_l^2}{\sum_{l=1}^{L} |\alpha_l|^2}-\left(\frac{\sum_{l=1}^{L} |\alpha_l|^2 \omega_l}{\sum_{l=1}^{L} |\alpha_l|^2}\right)^2}.
    \label{AS}
\end{align}
The measurement results are summarized in Table~\ref{tab_results}, and Fig.~\ref{fig_dsas} compares the empirical and fitted cumulative distribution functions (CDFs) of RMS DS and AS. Following~\cite{outdoor4}, five candidate distributions are fitted by maximum likelihood to the respective pooled samples of RMS DS and AS\footnote{Given the limited angular information provided by the four-element Rx array, parameter estimation is performed separately for the Tx-side multiple-input single-output (MISO) response at each Rx element. Accordingly, the angular analysis is restricted to the estimated AoDs \(\omega\), and the reported AS characterizes departure-side dispersion.}. The DS dataset contains 8192 element-pair samples from all 16 links, whereas the AS dataset contains 60 Rx-specific samples from TRx1--TRx15, including both LoS and OLoS links.

To compare these fits quantitatively, we use the two-sided KS distance, which measures the maximum absolute deviation between the empirical and fitted CDFs~\cite{kstest}. For samples \(x_1,\ldots,x_N\), the empirical CDF is \(\widehat F_N(x)=\frac{1}{N}\sum_{j=1}^{N}\mathbf{1}\{x_j\le x\}\), where \(\mathbf{1}\{\cdot\}\) is the indicator function. With \(F(x;\widehat{\boldsymbol\theta})\) denoting a continuous candidate CDF whose parameters are estimated from these samples, the KS distance is
\begin{align}
	D_{\mathrm{KS}}
	=\sup_{x\in\mathbb{R}}
	\left|\widehat F_N(x)-F(x;\widehat{\boldsymbol\theta})\right|
	=\max_{1\le i\le N}
	\left\{
	\begin{aligned}
		&\frac{i}{N}-F(x_{(i)};\widehat{\boldsymbol\theta}),\\
		&F(x_{(i)};\widehat{\boldsymbol\theta})-\frac{i-1}{N}
	\end{aligned}
	\right\},
	\label{eq_ks_distance}
\end{align}
where \(x_{(i)}\) denotes the \(i\)-th order statistic. A smaller \(D_{\mathrm{KS}}\) indicates closer CDF agreement, and the candidate with the smallest distance is preferred under this criterion. Table~\ref{tab_ks_fits} summarizes the fitted parameters and corresponding KS distances. Among the five candidates, LN provides the closest CDF agreement for both DS and AS, with \(D_{\mathrm{KS}}\) of \(0.0834\) and \(0.0683\), respectively. 

\begin{table*}
\centering
\caption{Fitted Distribution Parameters and KS Distances for DS, AS, and KF}
\label{tab_ks_fits}
\begingroup
\footnotesize
\setlength{\tabcolsep}{4pt}
\renewcommand{\arraystretch}{1.2}
\begin{tabular}{llcccccc}
\hline
\multirow{2}{*}{Distribution} & \multirow{2}{*}{Parameters}
& \multicolumn{2}{c}{DS (\(N=8192\))}
& \multicolumn{2}{c}{AS (\(N=60\))}
& \multicolumn{2}{c}{KF (\(N=8192\))} \\
\cline{3-8}
& & Estimates & \(D_{\mathrm{KS}}\) & Estimates & \(D_{\mathrm{KS}}\) & Estimates & \(D_{\mathrm{KS}}\) \\
\hline
Lognormal & \((\mu_{10},\sigma_{10})\) & \((-0.3477,0.1972)\) & \textbf{0.0834} & \((0.5532,0.2610)\) & \textbf{0.0683} & \((1.4479,0.4488)\) & \textbf{0.0453} \\
Gaussian & \((\mu,\sigma)\) & \((0.5001,0.2683)\) & 0.1508 & \((4.2984,2.9357)\) & 0.1862 & \((51.0704,90.4176)\) & 0.2977 \\
Nakagami & \((m,\Omega)\) & \((1.2066,0.3221)\) & 0.1358 & \((0.7905,27.0950)\) & 0.1582 & \((0.5000,10783.5271)\) & 0.3589 \\
Rician & \((\nu,\sigma)\) & \((0.0000,0.4013)\) & 0.1391 & \((0.0008,3.6807)\) & 0.1797 & \((0.0031,73.4286)\) & 0.5197 \\
Weibull & \((\lambda,k)\) & \((0.5648,1.9663)\) & 0.1381 & \((4.8434,1.6231)\) & 0.1283 & \((47.7399,0.8972)\) & 0.1090 \\
\hline
\end{tabular}

\endgroup
\end{table*}

For DS, the fitted parameters are \((\mu_{\mathrm{DS}},\sigma_{\mathrm{DS}})=(-0.348,0.197)\) in the \(\log_{10}(\mathrm{DS}/1~\mathrm{ns})\) domain. The DS values in Table~\ref{tab_results} range from \(0.227\) to \(1.023~\mathrm{ns}\), indicating limited delay dispersion in the measured responses. This is consistent with power concentrated near a dominant delay component and relatively weak contributions at larger excess delays. The observed delay structure also depends on the scattering geometry, antenna directivity, and path-extraction sensitivity.

For AS, the preferred LN model has parameters \((\mu_{\mathrm{AS}},\sigma_{\mathrm{AS}})=(0.553,0.261)\) in the \(\log_{10}(\mathrm{AS}/1^\circ)\) domain. Most samples in Fig.~\ref{fig_dsas}\subref{fig_as} are below \(8^\circ\), whereas the four Rx-specific estimates at TRx12 range from \(12.829^\circ\) to \(14.157^\circ\), with a mean of \(13.542^\circ\). These observations form the upper tail of the analyzed AS dataset and show that the large link-level value is not attributable to a single Rx estimate. The reported AS quantifies the departure-angle dispersion of the MPCs observed with the directional antennas.

All four examined neighboring LoS/OLoS pairs exhibit larger AS at the obstructed location. For example, AS increases from \(2.418^\circ\) at TRx5 to \(4.353^\circ\) at TRx6 and from \(4.287^\circ\) at TRx10 to \(5.651^\circ\) at TRx11. This contrast is consistent with vegetation-induced changes in angular power distribution. In LoS links, a strong direct component concentrates power around its departure direction. At THz frequencies, vegetation can attenuate this component through absorption and scattering, while limited diffraction hinders propagation around obstructions. When this component is more strongly attenuated than off-axis reflected and scattered MPCs, their relative contributions increase, broadening the observed departure-angle distribution. 


For context, the fitted logarithmic parameters are compared with the UMi street-canyon expressions in 3GPP TR~38.901~\cite{3gpp38901}. In the \(\log_{10}(\mathrm{DS}/1~\mathrm{s})\) domain, the measured parameter pair is \((-9.348,0.197)\), compared with extrapolated reference pairs of \((-7.749,0.380)\) for LoS and \((-7.439,0.686)\) for NLoS. The fitted AS parameters are also lower than the corresponding LoS and NLoS references, \((1.083,0.410)\) and \((0.946,0.609)\), respectively. These differences are consistent with frequency-dependent changes in the relative strengths of the observable MPCs.


\subsection{Rician K-Factor}

\begin{figure}
    \centering
    \includegraphics[width=3.4in]{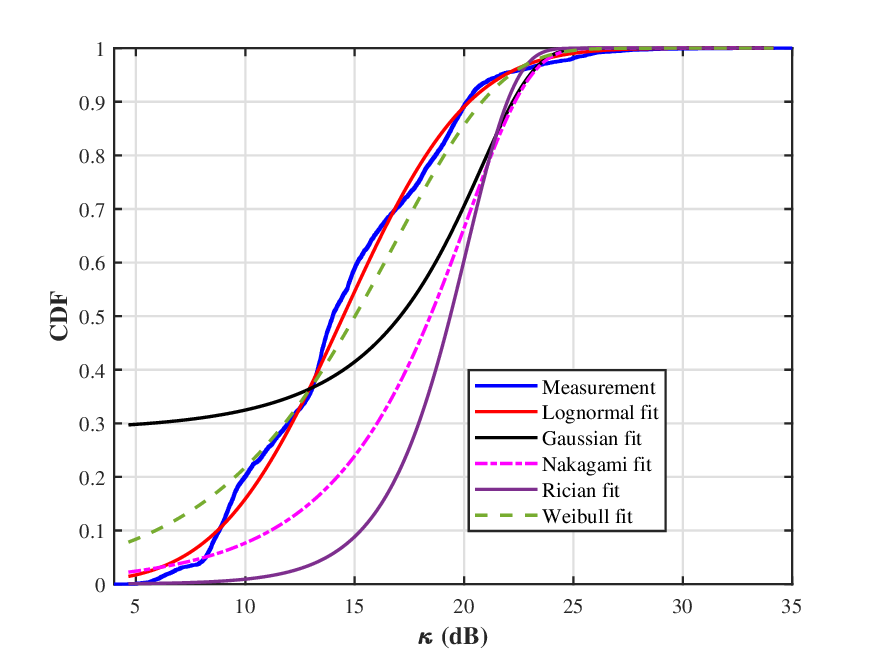}
    \caption{Empirical and fitted CDFs of the K-factor.}
\label{fig_kf}
\end{figure}

The Rician KF characterizes the power dominance of the strongest MPC relative to the remaining MPCs in the channel. It is defined as
\begin{align}
    \kappa
    =
    \frac{\max_{\tau}\mathrm{APDP}_{\tau}}
    {\sum_{\tau}\mathrm{APDP}_{\tau}-\max_{\tau}\mathrm{APDP}_{\tau}},
\end{align}
where \(\mathrm{APDP}_{\tau}\) denotes the aggregated APDP power of MPCs with closely spaced delays. Candidate distributions are fitted to the linear KF ratio \(\kappa\) using 8192 samples pooled from the 16 TRx links. Fig.~\ref{fig_kf} displays the empirical and fitted CDFs against \(10\log_{10}\kappa\) in dB. Table~\ref{tab_ks_fits} reports the parameter estimates and KS distances computed using Eq.~\eqref{eq_ks_distance}. Among the candidates, LN yields the smallest KS distance, \(D_{\mathrm{KS}}=0.0453\).

The LN fit to the pooled linear KF samples corresponds to a mean of \(\mu_{\kappa}=14.48~\mathrm{dB}\) and a standard deviation of \(\sigma_{\kappa}=4.49~\mathrm{dB}\) for \(10\log_{10}\kappa\), indicating pronounced power dominance of the strongest delay cluster over the remaining multipath. The mean exceeds the 3GPP UMi street-canyon LoS reference of \(9~\mathrm{dB}\), whereas the standard deviation is comparable to the reference value of \(5~\mathrm{dB}\)~\cite{3gpp38901}. 

The location-specific KF fluctuations provide a complementary view of the array response. TRx1 has a relatively large \(\sigma_\kappa\), indicating substantial variation in the dominant-to-residual power ratio across the sampled element pairs. Its near-field behavior is assessed separately in Section~\ref{nearfield}. The OLoS locations also have larger \(\sigma_\kappa\) than their neighboring LoS references in the TRx6/TRx5, TRx7/TRx8, TRx11/TRx10, and TRx13/TRx14 comparisons. These fluctuations are consistent with changes in path visibility and relative component strength as the antenna position varies near vegetation. 

\subsection{Correlation Properties}

\begin{figure*}
	\centering
	\subfloat[LoS.]{
		\includegraphics[width=0.48\textwidth]{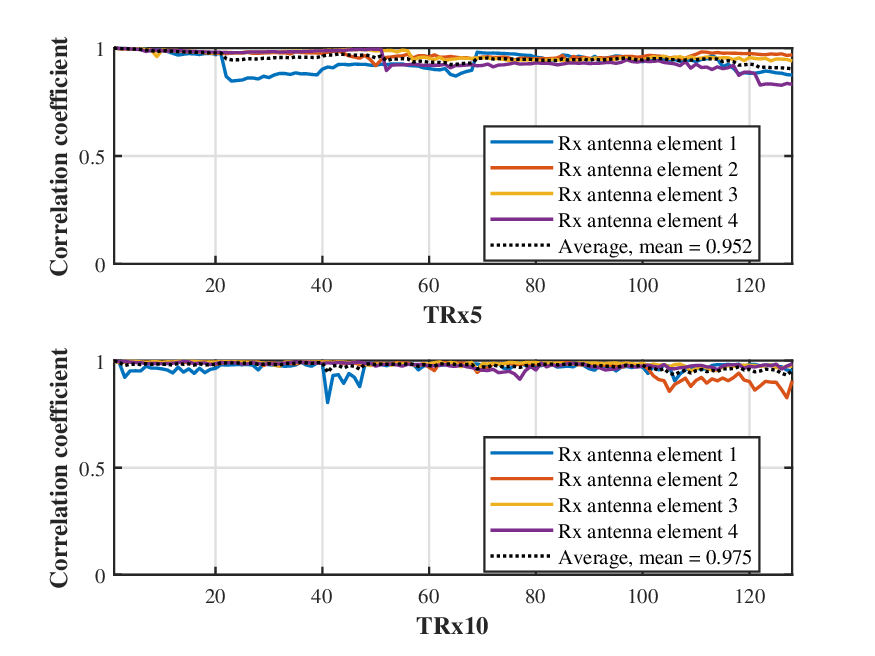}
		\label{fig_cc1}
	}
	\hfill
	\subfloat[OLoS.]{
		\includegraphics[width=0.48\textwidth]{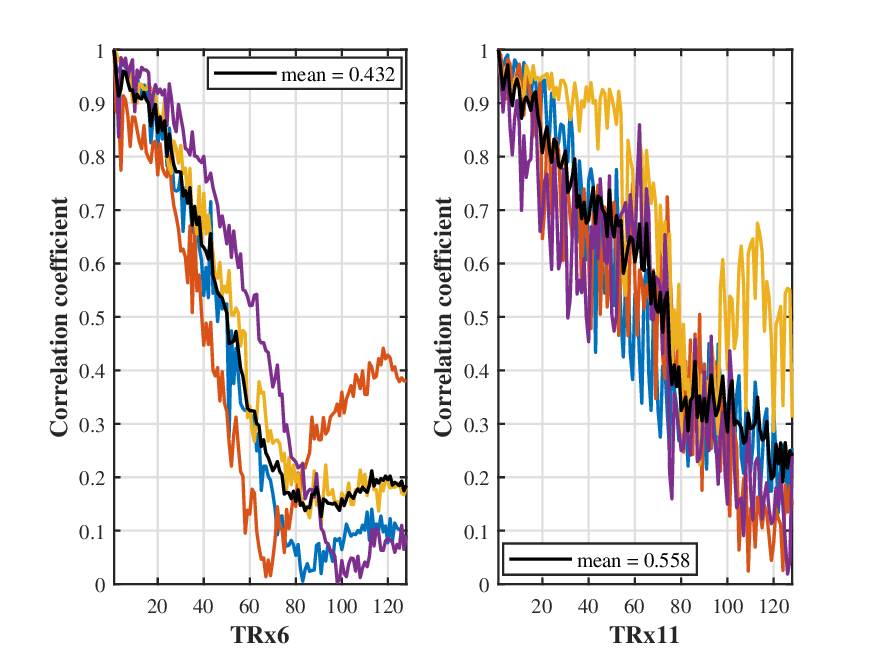}
		\label{fig_cc2}
	}
	\caption{CIR correlation coefficients referenced to the first Tx antenna element at
	(a) LoS locations TRx5 and TRx10 and
	(b) OLoS locations TRx6 and TRx11.}
	\label{fig_cc}
\end{figure*}

The CIR correlation coefficient between the \((r,m)\)-th CIR \(\mathbf{c}^{(r,m)}\) and \((r,n)\)-th CIR \(\mathbf{c}^{(r,n)}\) is defined as
\begin{align}
	\rho_{m,n}^{(r)}
	=
	\frac{
		\mathbb{E}
		\left[
		\left|
		\left(
		\mathbf{c}^{(r,m)}
		-
		\bar{\mathbf{c}}^{(r,m)}
		\right)^{\mathrm{H}}
		\left(
		\mathbf{c}^{(r,n)}
		-
		\bar{\mathbf{c}}^{(r,n)}
		\right)
		\right|
		\right]
	}{
		\sqrt{
			\mathbb{E}
			\left[
			\|
			\mathbf{c}^{(r,m)}
			-
			\bar{\mathbf{c}}^{(r,m)}
			\|_2^2
			\right]
			\mathbb{E}
			\left[
			\|
			\mathbf{c}^{(r,n)}
			-
			\bar{\mathbf{c}}^{(r,n)}
			\|_2^2
			\right]
		}
	},
	\label{corr}
\end{align}
where \(\bar{\mathbf{c}}^{(r,m)}\) and \(\bar{\mathbf{c}}^{(r,n)}\) denote the mean components of the corresponding CIRs. Fig.~\ref{fig_cc} shows the four Rx-specific correlation curves and their pointwise average across the four Rx elements. The link-level mean values reported below exclude the reference element, whose self-correlation is unity.

The two LoS examples maintain high correlation across the Tx aperture. Their Rx-averaged coefficients remain above 0.90, with link-level means of 0.952 at TRx5 and 0.975 at TRx10. This behavior indicates similar relative delay-domain structures across Tx positions, consistent with a persistent dominant direct component. The corresponding means are lower at the neighboring OLoS locations: 0.432 at TRx6 and 0.558 at TRx11. In both OLoS examples, the Rx-averaged correlation generally decreases with separation from the reference element, with local fluctuations, and reaches 0.184 and 0.240 at \(n=N_T\), respectively. The lower correlations at the OLoS locations are consistent with vegetation obstruction weakening the dominant component and increasing the sensitivity of the CIR to spatial variations in the relative contributions of reflected and scattered paths. Together with the larger AS values and stronger KF fluctuations discussed above, the consistent contrast across both location pairs supports stronger spatial variability of the CIR structure under obstruction.
\subsection{MIMO Properties}

\subsubsection{Near-Field Effects}\label{nearfield}
\begin{figure*}
	\centering
	\begin{minipage}[t]{0.48\textwidth}
		\centering
		\includegraphics[width=\linewidth]{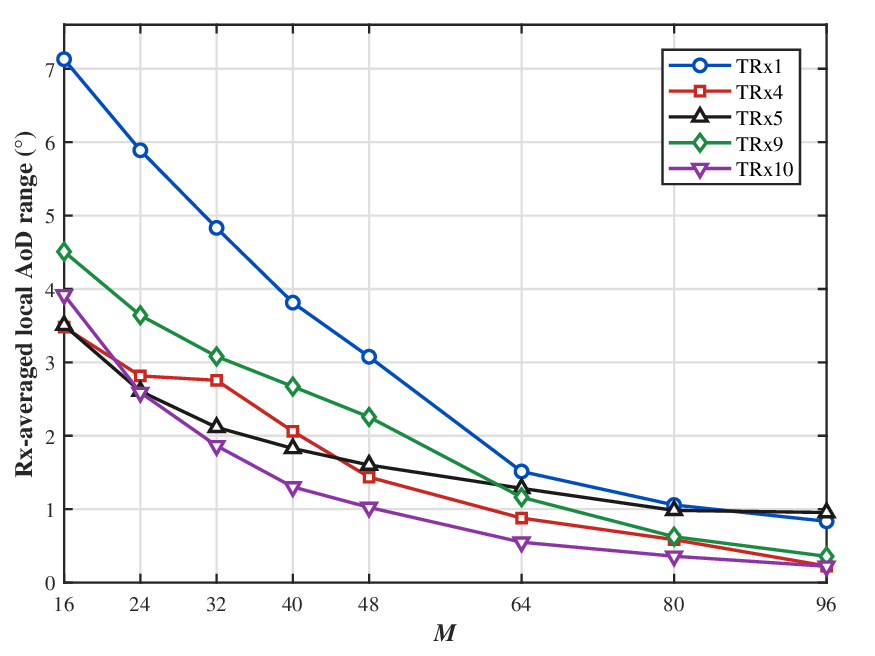}
		\caption{Rx-averaged local AoD range versus subarray size \(M\).}
		\label{fig_nf_size}
	\end{minipage}
	\hfill
	\begin{minipage}[t]{0.48\textwidth}
		\centering
		\includegraphics[width=\linewidth]{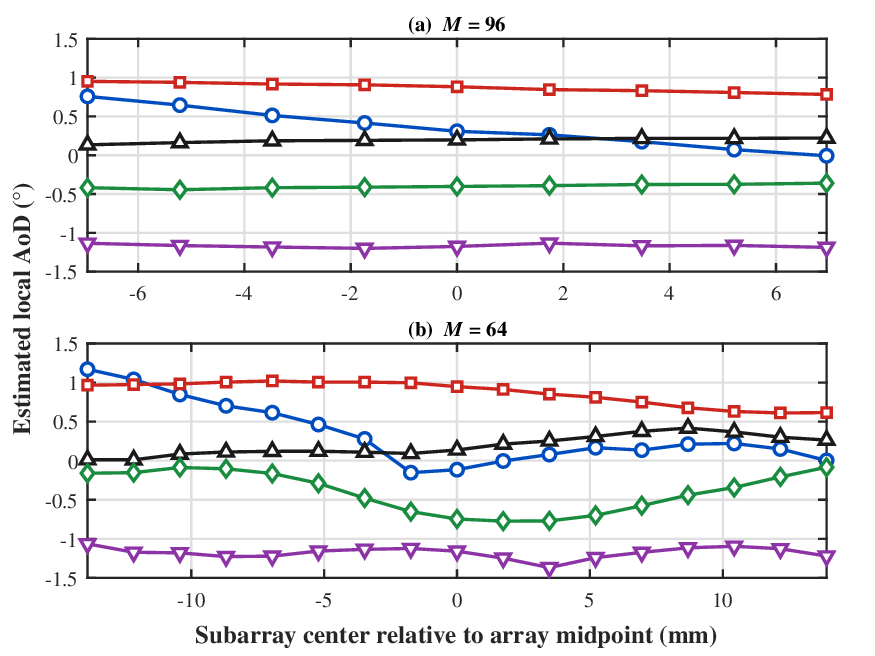}
		\caption{Local AoD estimates versus subarray-center position
			for (a) \(M=96\) and (b) \(M=64\).}
		\label{fig_nf_aod}
	\end{minipage}
\end{figure*}

Spherical wavefront curvature produces position-dependent local AoDs across the Tx aperture. At \(f_c=345~\mathrm{GHz}\), the wavelength \(\lambda_c\approx0.869~\mathrm{mm}\) and full Tx aperture \(D=(N_T-1)d_T\approx55.18~\mathrm{mm}\) yield a conventional broadside Rayleigh distance of \(2D^2/\lambda_c\approx7.01~\mathrm{m}\)~\cite{VAA}. Under this criterion, TRx1 lies in the radiative near field, whereas TRx4, TRx5, TRx9, and TRx10 serve as far-field LoS references. Local AoD variations are examined to complement this geometric assessment.

Following the sliding-subarray approach in~\cite{SAGE3}, the Tx-side MISO response at each Rx element is processed independently using the PD-aware SAGE procedure in Section~\ref{PDSAGE}. Each window contains \(M\) consecutive Tx elements and advances with a four-element stride. For \(M=96\) and \(64\), the subarray Rayleigh distances are approximately \(3.92\) and \(1.72~\mathrm{m}\), respectively, supporting a local plane-wave approximation at TRx1. For each Rx, the local AoD range is defined as the maximum minus the minimum AoD over all available windows, with the strongest component selected independently in each window. This range is then averaged over the four Rx elements.

As shown in Fig.~\ref{fig_nf_size}, the mean AoD range at TRx1 increases from \(0.834^\circ\) for \(M=96\) to \(7.131^\circ\) for \(M=16\), compared with \(3.480^\circ\)--\(4.509^\circ\) for the far-field references at \(M=16\). TRx1 exhibits the largest mean range for the tested sizes \(M=16\)--80. Reducing \(M\) decreases the local estimation aperture and increases the span of the window centers, \((N_T-M)d_T\). The observed size dependence therefore reflects the combined effects of aperture, sliding coverage, and estimation variability. Comparisons at a fixed \(M\) provide a more direct assessment of the interlink angular contrast.

Fig.~\ref{fig_nf_aod} compares the local AoD trajectories at selected Rx elements. For \(M=96\), the estimated local AoD at the selected Rx element of TRx1 varies approximately monotonically over a range of \(0.765^\circ\), compared with \(0.068^\circ\)--\(0.169^\circ\) for the far-field references. For \(M=64\), the TRx1 range increases to \(1.324^\circ\), compared with \(0.304^\circ\)--\(0.690^\circ\) for the far-field references, and its trajectory exhibits more pronounced local fluctuations. Together with the geometric assessment, the observed angular contrast is consistent with more pronounced near-field effects at TRx1 than at the longer-range links.

\begin{figure}[!t]
	\centering
	
	\subfloat[TRx1.]{
		\includegraphics[width=0.97\columnwidth]{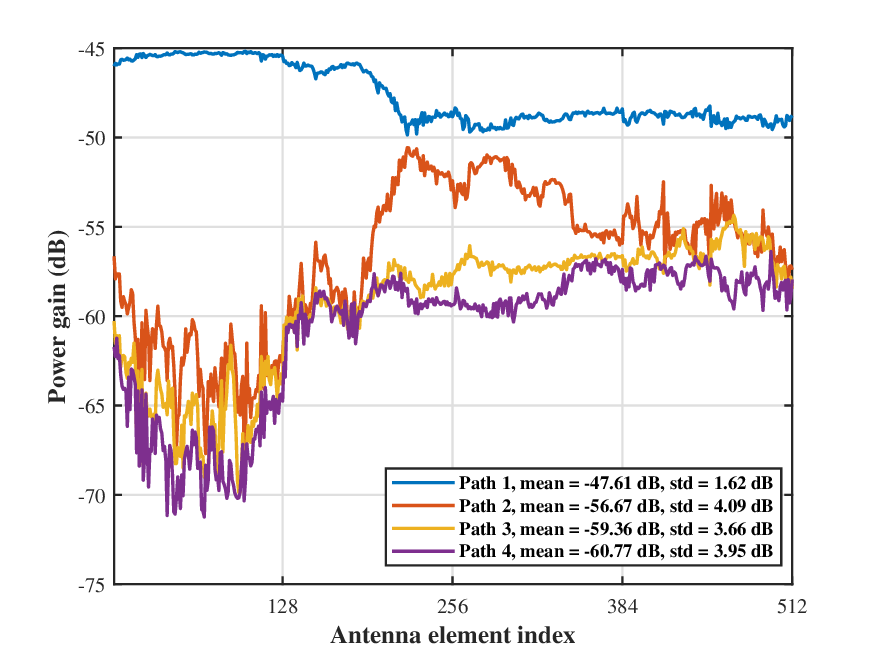}
		\label{fig_sns1}
	}
	
	\vspace{1mm}
	
	\subfloat[TRx6.]{
		\includegraphics[width=0.97\columnwidth]{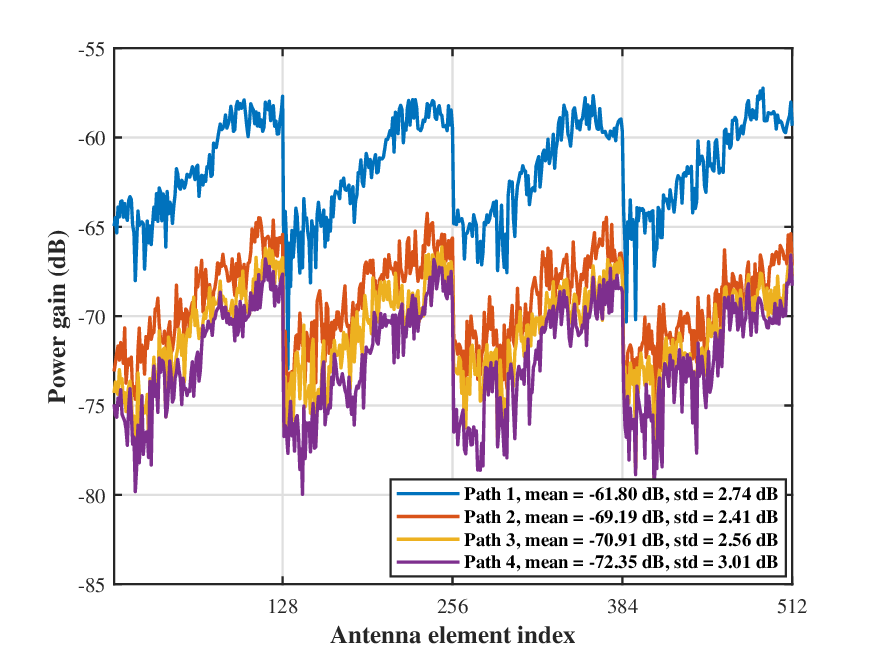}
		\label{fig_sns2}
	}
	
	\vspace{1mm}
	
	\subfloat[TRx10.]{
		\includegraphics[width=0.97\columnwidth]{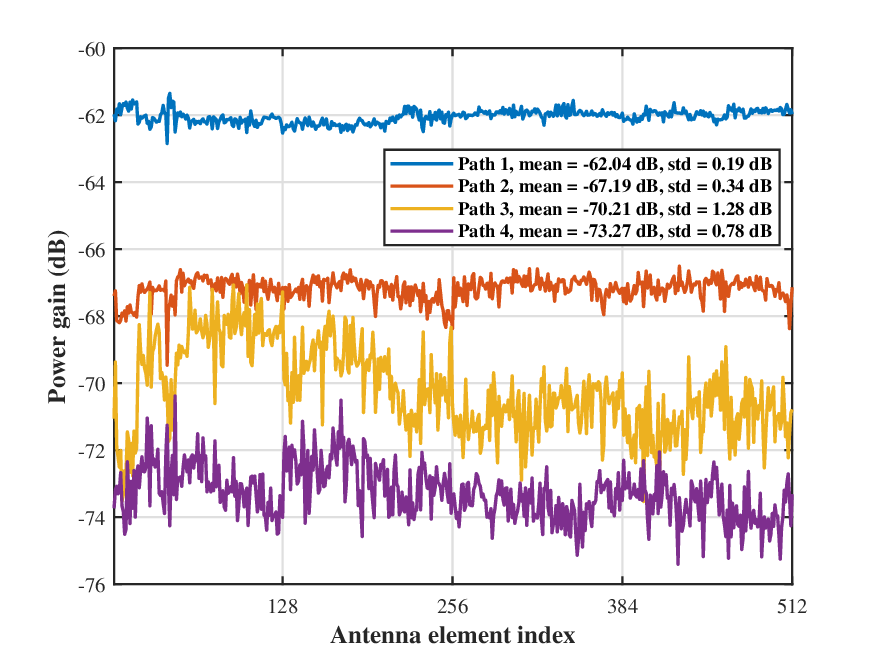}
		\label{fig_sns3}
	}
	
	\caption{Power gain profiles across Tx--Rx element pairs
	for the four paths with the highest average power at
	(a) TRx1 (LoS, near-field), (b) TRx6 (OLoS, far-field), and (c) TRx10 (LoS, far-field).}
	\label{fig_sns}
\end{figure}

\begin{figure}[!t]
	\centering
	\subfloat[TRx1.]{
		\includegraphics[width=0.97\columnwidth]{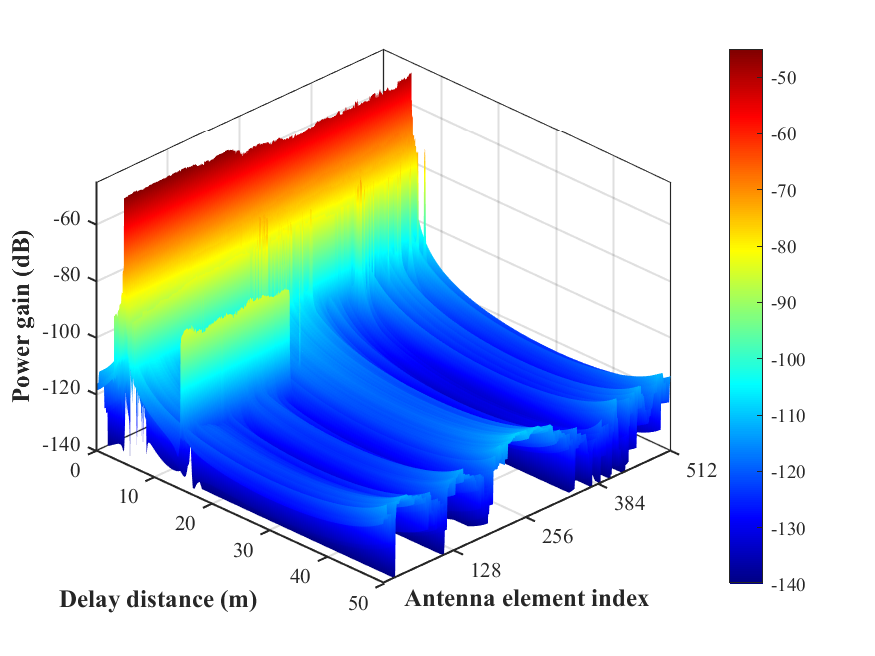}
		\label{fig_clusterBD1}
	}
	\vspace{1mm}
	\subfloat[TRx4.]{
		\includegraphics[width=0.97\columnwidth]{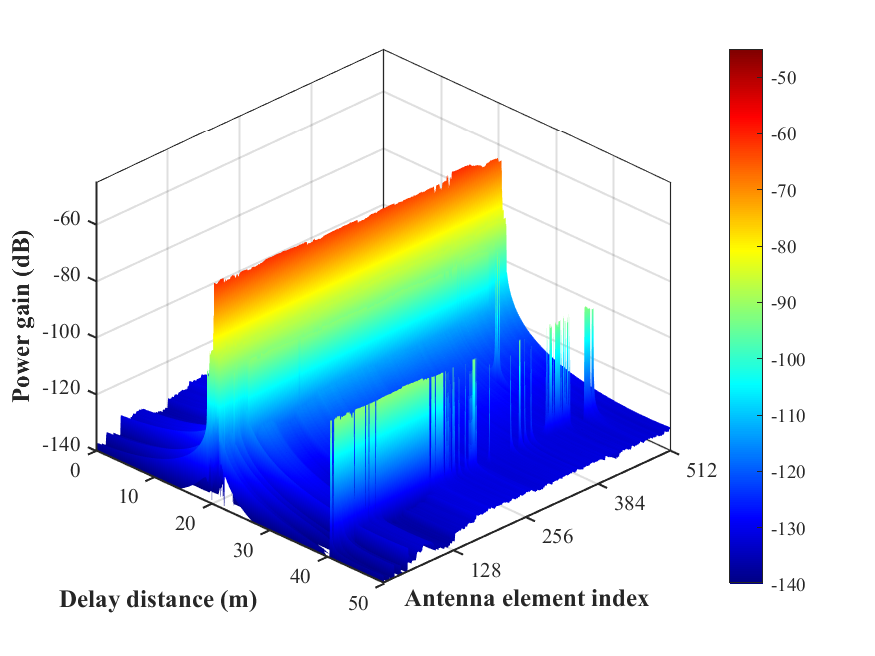}
		\label{fig_clusterBD2}
	}
	\caption{PDPs across Tx--Rx element pairs at
		(a) TRx1 and (b) TRx4.}
	\label{fig_clusterBD}
\end{figure}

\subsubsection{Spatial Non-stationary Characteristics}

Fig.~\ref{fig_sns} compares the power profiles of the four extracted paths with the highest average power at TRx1, TRx6, and TRx10. The horizontal axis indexes 512 Tx--Rx element pairs, arranged as four Rx blocks of 128 Tx positions. The reported means and standard deviations are evaluated in the dB domain over these pairs, characterizing the power levels and variations of individual components.

At TRx1, the path powers vary appreciably across the sampled element pairs, with Path~2 exhibiting a standard deviation of \(4.09~\mathrm{dB}\). Antenna directivity and changes in path visibility may contribute to these variations~\cite{sns2}. In contrast, TRx10 exhibits comparatively stable path powers, with standard deviations ranging from \(0.19\) to \(1.28~\mathrm{dB}\) and a nearly constant dominant component. The path-power fluctuations at the OLoS location TRx6 are greater than those at TRx10, with standard deviations
of \(2.41\)--\(3.01~\mathrm{dB}\). Similar Tx-dependent profiles recur across the four Rx blocks. This behavior is consistent with the closely spaced Rx elements observing a similar obstruction geometry, while path visibility and relative strength vary along the Tx aperture. These observations indicate that the extent of spatial power variation differs among links and individual paths.

\subsubsection{Cluster Birth--Death Property}

Fig.~\ref{fig_clusterBD} presents the PDPs across the sampled Tx--Rx element pairs at TRx1 and TRx4. The components at delay distances of approximately \(15~\mathrm{m}\) at TRx1 and \(40~\mathrm{m}\) at TRx4 form ridge-like segments over limited index intervals. At TRx1, the reflected component is mainly visible in the Rx1 block and is substantially weaker or not evident in the remaining blocks. These observations indicate that the corresponding propagation contributions are spatially localized within the sampled array.

Such localized contributions are consistent with a visibility-region (VR) description, in which a cluster is observable over a limited subset of the sampled antenna pairs. Transitions in cluster visibility are represented as array-domain birth and death, although the observed boundaries may also depend on measurement sensitivity and extraction settings. These observations motivate incorporating spatially varying cluster visibility into THz-band MIMO channel models~\cite{sns}.

\section{Conclusion}\label{conclusion}

In this paper, we conducted a 330--360~GHz channel measurement campaign in a rectangular outdoor street area at PML using a \(128\times4\) VAA-based MIMO configuration. The environmental stationarity was verified, and a PD-aware SAGE algorithm was proposed to improve delay resolution and channel parameter estimation accuracy
during measurement post-processing. Based on the processed data, the channel characteristics of LoS and OLoS links were analyzed and modeled. The results showed that outdoor THz-band propagation exhibits characteristics distinct from those observed at mmWave frequencies, particularly in terms of PL, SF, DS, AS, KF, and correlation properties. Statistical fitting was further performed for key channel parameters, and the KS distance was used to compare candidate distributions, favoring LN models for DS, AS, and KF within the analyzed dataset. MIMO-specific features, including near-field effects, SnS and cluster birth--death property, were also investigated. The comparative analysis between LoS and OLoS links revealed the impact of vegetation obstruction on outdoor THz propagation characteristics.

\bibliographystyle{IEEEtran}
\bibliography{ref}



\end{document}